\documentclass[fleqn,usenatbib]{mnras}
\usepackage{amssymb,amsmath}
\usepackage{fontawesome}
\usepackage{lipsum}
\usepackage{pifont}
\usepackage{bm}
\usepackage{ulem}
\usepackage{balance}
\usepackage{tabularx}
\usepackage{multirow}
\usepackage{mathtools}
\usepackage{enumitem}
\usepackage{booktabs}
\usepackage{appendix}
\usepackage{listings}

\newcommand{\RT}{$R_\textup{200c}$}
\newcommand{\YT}{$Y_\textup{200c}$}

\newcommand{\Msg}{$M_*/M_\mathrm{gas}$}

\newcommand{\be}{\begin{equation}}
\newcommand{\ee}{\end{equation}}
\def\Mpc{\, h^{-1} \, {\rm Mpc}}

\def\Ms{\, h^{-1} \, M_\odot}
\defcitealias{VilAngGen20}{V21}
\defcitealias{Wad22}{W22}
\defcitealias{Pan21}{P20}

\title[$Y$-$M$ for low-mass halos]{The SZ flux–mass ($Y$-$M$) relation at low halo masses: improvements with symbolic regression and\\ strong constraints on baryonic feedback}

 \author[Wadekar et al.]{Digvijay Wadekar,$^{1}$\thanks{E-mail: jayw@ias.edu (DW)}
Leander Thiele,$^{2}$
J.~Colin Hill,$^{3,4}$
Shivam Pandey,$^{5}$\newauthor
Francisco Villaescusa-Navarro,$^{4,6}$
David N. Spergel,$^{4,6}$
Miles Cranmer,$^{6}$
Daisuke Nagai,$^{7}$\newauthor
Daniel Angl\'es-Alc\'azar,$^{8,4}$
Shirley Ho,$^{4,6,9}$
Lars Hernquist$^{10}$\\
\\
$^{1}$School of Natural Sciences, Institute for Advanced Study, 1 Einstein Drive, Princeton, NJ 08540, USA\\
$^{2}$Department of Physics, Princeton University, Jadwin Hall, Princeton NJ 08544, USA\\
$^{3}$Department of Physics, Columbia University, New York, NY 10027, USA\\
$^{4}$Center for Computational Astrophysics, Flatiron Institute, 162 5th Ave, New York, NY 10010, USA\\
$^{5}$Department of Physics and Astronomy, University of Pennsylvania, Philadelphia, PA 19104, USA\\
$^{6}$Department of Astrophysical Sciences, Princeton University, Peyton Hall, Princeton NJ 08544-0010, USA\\
$^{7}$Department of Physics, Yale University, New Haven, CT 06520, USA\\
$^{8}$Department of Physics, University of Connecticut, 196 Auditorium Road, Storrs, CT, 06269, USA\\
$^{9}$Department of Physics, Carnegie Mellon University, Pittsburgh, PA 15217, USA\\
$^{10}$Center for Astrophysics | Harvard \& Smithsonian, 60 Garden Street, Cambridge, MA 02138, USA\\
}

\date{Accepted XXX. Received YYY; in original form ZZZ}

\pubyear{2021}

\begin{document}
\label{firstpage}
\pagerange{\pageref{firstpage}--\pageref{lastpage}}
\maketitle

\begin{abstract}
Feedback from active galactic nuclei (AGN) and supernovae can affect measurements of integrated SZ flux of halos ($Y_\mathrm{SZ}$) from CMB surveys, and cause its relation with the halo mass ($Y_\mathrm{SZ}-M$) to deviate from the self-similar power-law prediction of the virial theorem. We perform a comprehensive study of such deviations using CAMELS, a suite of hydrodynamic simulations with extensive variations in feedback prescriptions. 
We use a combination of two machine learning tools (random forest and symbolic regression) to search for analogues of the $Y-M$ relation which are more robust to feedback processes for low masses ($M\lesssim 10^{14}\, h^{-1} \, M_\odot$); we find that simply replacing $Y\rightarrow Y(1+M_*/M_\mathrm{gas})$ in the relation makes it remarkably self-similar. This could serve as a robust multiwavelength mass proxy for low-mass clusters and galaxy groups. Our methodology can also be generally useful to improve the domain of validity of other astrophysical scaling relations.

We also forecast that measurements of the $Y-M$ relation could provide percent-level constraints on certain combinations of feedback parameters and/or rule out a major part of the parameter space of supernova and AGN feedback models used in current state-of-the-art hydrodynamic simulations. Our results can be useful for using upcoming SZ surveys (e.g.~SO, CMB-S4) and galaxy surveys (e.g.~DESI and Rubin) to constrain the nature of baryonic feedback. Finally, we find that the an alternative relation, $Y-M_*$, provides complementary information on feedback than $Y-M$. \href{https://github.com/JayWadekar/ScalingRelations_ML}{\faGithub}


\end{abstract}

\begin{keywords}
cosmology: observations; cosmic background radiation; large-scale structure of Universe; galaxies: clusters: general; galaxies: groups: general
\end{keywords}

\section{Introduction}
\label{sec:intro}

Baryonic feedback is one of the leading sources of systematic uncertainty in many areas of cosmological inference. 
One way of dealing with the uncertainty is to remove the small scales affected by feedback from the analysis (see e.g., weak lensing survey analyses \citep{Sec22, Amo22, Sem11}); another way is to introduce parameters which encode the feedback effects and later marginalize over them. Both of these approaches ultimately weaken the cosmological constraints (see e.g., Fig.~1 of \cite{WadIva20} corresponding to the BOSS survey). Hydrodynamic simulations can in principle be used to model the baryonic effects, however different simulations currently give different results based on the sub-grid prescriptions used in them \citep{Chi18, Van14,Som15, Har15, VilAngGen20,Vog20}.
It is therefore imperative to either narrow down the uncertainty in the feedback prescriptions, or to make the cosmological inference pipelines more robust to feedback effects. We explore both of these paths in the context of the Sunyaev-Zeldovich (SZ) scaling relations for galaxy groups and clusters in this paper.

We focus on the self-similar power-law relationship between the integrated electron pressure ($Y_\mathrm{SZ}$, hereafter $Y$) and the mass of the halo (M), known as the $Y-M$ relation: $M\propto Y^{3/5}_\mathrm{SZ}$. This relation approximately follows from the virial theorem \citep{Kai86,Bry98,KraBor12}. $Y-M$ has been crucial for cosmological analysis of clusters as it has been widely used to infer the masses of clusters from measurements of $Y_\mathrm{SZ}$ in CMB surveys like Planck, ACT or SPT\footnote{It is worth mentioning that in cluster cosmology analyses, the normalization and slope of the $Y-M$ relation is calibrated directly from data (e.g., using weak-lensing-inferred halo masses). The fitted values of the power-law exponent are however fairly close to the 3/5 prediction from the virial theorem, see e.g., \cite{BatBon12}.} \citep{Ade13,Ade15,Ade15b,Has13,Hil21,Boc15,Boc19}.

However, the virial theorem is based on the assumption that the only source of energy input into the intra-cluster medium is gravitational. For high-mass clusters ($M_{200c}\gtrsim10^{14} \Ms$) which have deep potential wells, the power-law relation is fairly accurate as gas ejected from various feedback processes cannot escape the virial radius \citep{Kra05}. This is however not true for low-mass clusters or groups as they have much shallower potential wells. Feedback from active galactic nuclei (AGN) can eject gas from these systems, while supernova feedback (SN) can also significantly change the gas thermal energy. Therefore the $Y-M$ relationship can deviate from self-similarity for low-mass halos (also referred to as a break in the power-law relation). There have been multiple recent cosmological analyses hinting at a break \citep{Hil18,Gre15, Gat21,Pan21, Osa18, Sin21}. Furthermore, such breaks have also been reported in multiple recent hydrodynamic simulations; e.g., COSMO-OWLS \citep{LeB17,LeB15}, IllustrisTNG \citep{Pop21,Wad22, Lee22}, SIMBA \citep{Yan22,RobDav20,RobDav21}.

Our goal in this paper is two-fold:
$(i)$ make the $Y-M$ relation more robust to feedback from baryonic processes so that it could be used in mass estimation of low-mass clusters and galaxy groups,
$(ii)$ determine the constraining power of measurements of $Y-M$ on baryonic feedback models used in hydrodynamic simulations.

For the first goal, we use machine learning tools to model the following function $f$
\be
\frac{M^{5/3}} {Y}= f(\{i_\textup{obs}\})
\label{eq:MLintro}
\ee
where $\{i_\textup{obs}\}$ is the set of various observable properties of groups/clusters from multi-wavelength surveys (e.g., gas mass, gas temperature or density profile, gas anisotropy, richness, galaxy colors/magnitudes, half-light radius, among many more).
Such a modeling can enable us to find a new relationship: $Y\, f(\{i_\textup{obs}\}) - M$ which has a smaller deviation from self-similarity as compared to $Y-M$ (in other words, we want to find a function which anti-correlates with the deviation from self-similarity). This problem is challenging for manual data analysis methods because the set of $\{i_\textup{obs}\}$ is high-dimensional and $f$ can be a non-linear function. Furthermore, there are no obvious first principles predictions for which properties in $\{i_\textup{obs}\}$ should contribute because groups and clusters are non-linear objects.

Note that in our companion paper, \cite{Wad22}, we also tried to improve $Y-M$ using machine learning tools. 
However, we only focused on high mass halos (corresponding to large clusters: $M_{200c}\gtrsim 10^{14}\Ms$), where the traditional $Y-M$ relation is already close to a power-law. We had found an alternative relationship $Y(1\,-\, A\, c_\mathrm{gas})-M$ has a lower scatter in the high-mass regime than $Y-M$ ($c_\mathrm{gas}$ is the concentration of ionized gas within the cluster).
In this paper, we instead focus on the small clusters or group mass regime ($M_{200c}\lesssim 10^{13.5}\Ms$), where $Y-M$ significantly deviates from a power-law. We found that directly using the previous relationship did not lead to an appreciable improvement for low masses, and hence we look for an alternative relationship for this case with the form of Eq.~\ref{eq:MLintro}.
To ensure that the alternative relationship is robust to feedback, we will use the CAMELS suite of simulations (which contains 2,184 hydrodynamic simulations run with extensive variations of baryonic feedback parameters employing two independent codes \citep{VilAngGen20}). It is worth noting that the high-mass clusters are used for cosmological measurements and therefore the masses are required to be accurately measured to within $\sim$10\%; on the other hand, inferring halo masses of groups is useful for astrophysical applications and an accuracy of roughly a factor of $\sim$2 in their masses can be sufficient.

For our second goal of constraining baryonic feedback with $Y-M$, we use the fact that while $Y$ is measured in SZ surveys, the halo mass ($M$) can be inferred in various ways with galaxy surveys: e.g., using shear or galaxy maps \citep{Gat21,Pan21,Aga11,Ade13,Hil18,Han11,Gre15,Jim18}. As different hydrodynamic simulations give different predictions for $Y-M$ for low-masses, one could therefore attempt to use the observations of $Y-M$ to constrain the feedback prescriptions in the simulations (see e.g., \cite{LeB15, LeB17, Hil18,Pan20, Pan21,Yan22}). We will use a fisher matrix formalism and use CAMELS to obtain derivatives of $Y-M$ as a function of parameters corresponding to the strength of supernova or AGN feedback. We will then derive constraints on these parameters by using forecasted errorbars on $Y-M$ from upcoming CMB and galaxy surveys.
It is also worth mentioning that, apart from $Y-M$, various other SZ-related measurements from current and upcoming surveys can also be used as probes of baryonic feedback \citep{Amo21,Sch21,Soe17,Hal19,Pan21,Nic22, Mos22,Cha08,Cha07,Lac19,Dut17,Rua15}.

%

The remainder of the paper is organized as follows. In Section~\ref{sec:ClusterData}, we briefly describe the cluster data that we use from various hydrodynamical simulations. In Section~\ref{sec:YM}, we present an overview of the $Y-M$ relation. We discuss an overview of our ML techniques in Section~\ref{sec:ML}. We present results for improving the $Y-M$ relation in Section~\ref{sec:Results1}. The results for constraining baryonic feedback are in Section~\ref{sec:Results2}. We conclude and discuss possible future work in Section~\ref{sec:Conclusions}.

\begin{table*}
    \centering
    \begin{tabular}{|p{1.5cm}||p{2.5cm}|p{2cm}|p{4.5cm}|p{3.5cm}|}
    \hline
    \hline
    Simulation & {\rm SN1} & {\rm SN2} & {\rm AGN1} &  {\rm AGN2}\\
    \hline
    \hline
 IllustrisTNG & Galactic winds: & Galactic winds: & Kinetic mode BH feedback: & Kinetic mode BH feedback: \\[0.5ex]
& energy per unit SFR & wind speed & energy per unit BH accretion rate  & ejection speed / burstiness\\
    \hline
SIMBA & Galactic winds: & Galactic winds: & QSO \& jet-mode BH feedback: & Jet-mode BH feedback:\\[0.5ex] 
& mass loading & wind speed & momentum flux & jet speed\\
    \hline
    \end{tabular}
\caption{
A brief description of the physical meaning of the four astrophysical parameters which are varied in the IllustrisTNG and SIMBA suites of CAMELS (SN [AGN] corresponds to feedback from supernova [active galactic nuclei]). This table is taken from \citet{VilAngGen20} and is provided here for self-contained discussion.}
\label{table:t1}
\end{table*}

\section{Data and properties of clusters and groups}
\label{sec:ClusterData}

In this section, we provide a brief description of the cluster data that we use in our analysis.
We primarily use the CAMELS suite of simulations\footnote{CAMELS: \url{https://www.camel-simulations.org}} \citep{VilAngGen20} 
which consists of 2,184 hydrodynamic simulations 
(each simulation box has side-length 25 $\Mpc$).
We outline some important properties of CAMELS in the following and refer the reader to \cite{VilAngGen20, CAMELS_public} for further details. The CAMELS boxes are simulated with variations in two cosmological parameters ($\Omega_m$ and $\sigma_8$) and four parameters corresponding to feedback from supernovae (SN1, SN2) and active galactic nuclei (AGN1, AGN2). 
Broadly, the SN parameters encode the subgrid prescription for galactic winds,
while the AGN parameters describe the efficiency of black hole feedback (see Table~\ref{table:t1} for details).

\begin{figure*}
\centering
\includegraphics[scale=0.5,keepaspectratio=true]{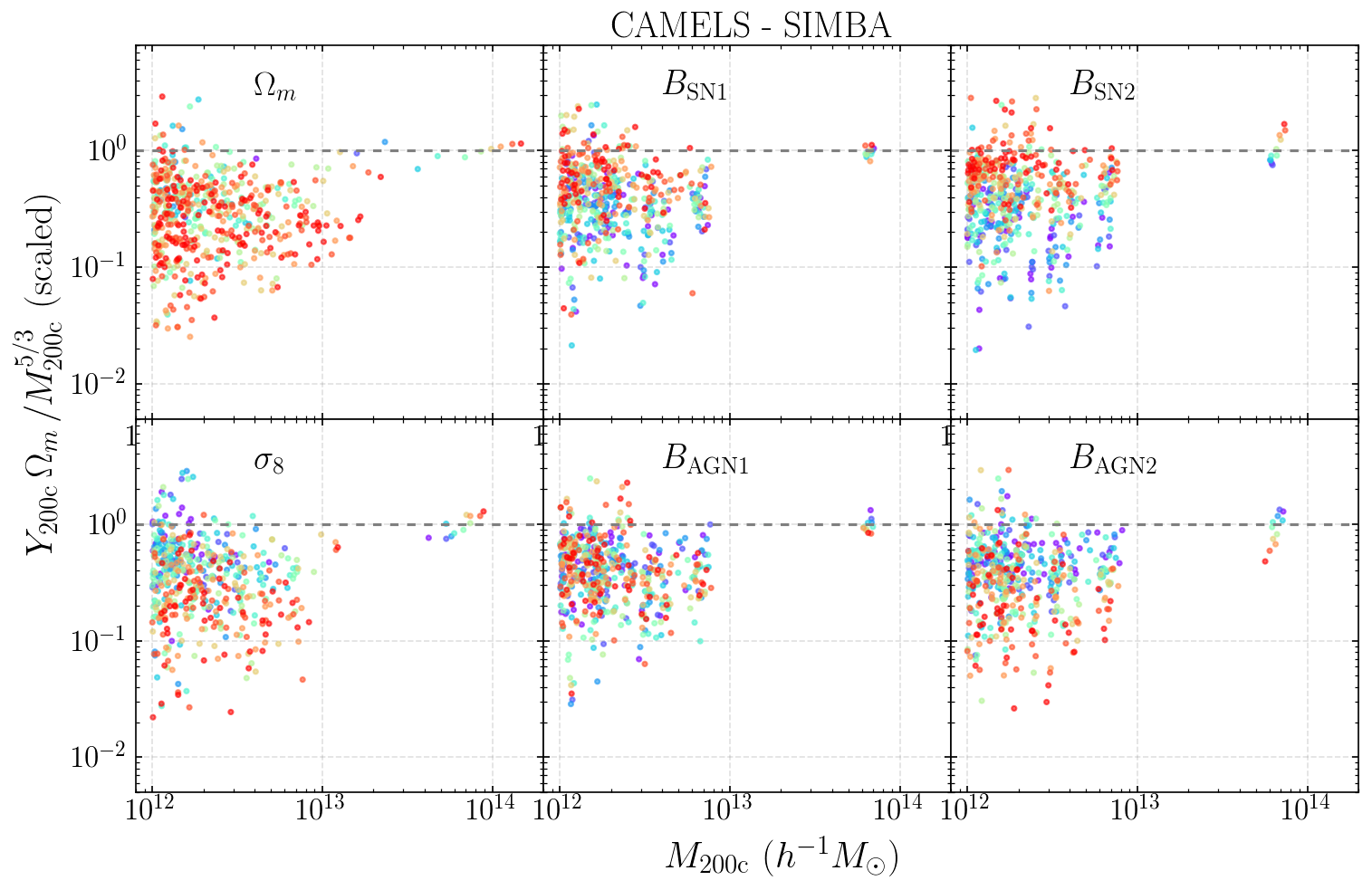}
\includegraphics[scale=0.25,keepaspectratio=true]{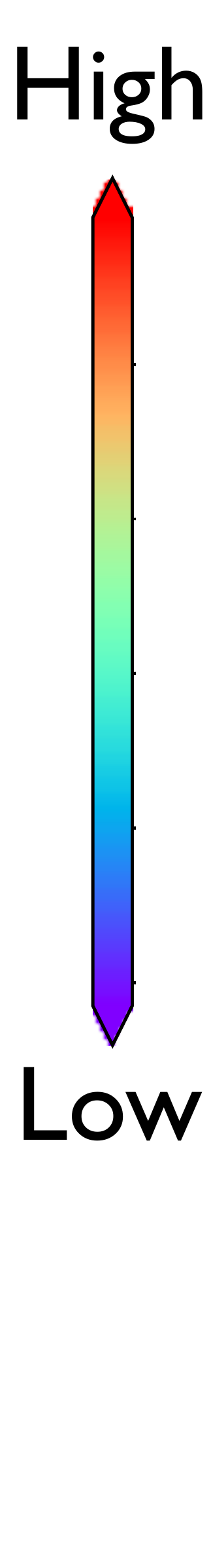}
\includegraphics[scale=0.5,keepaspectratio=true]{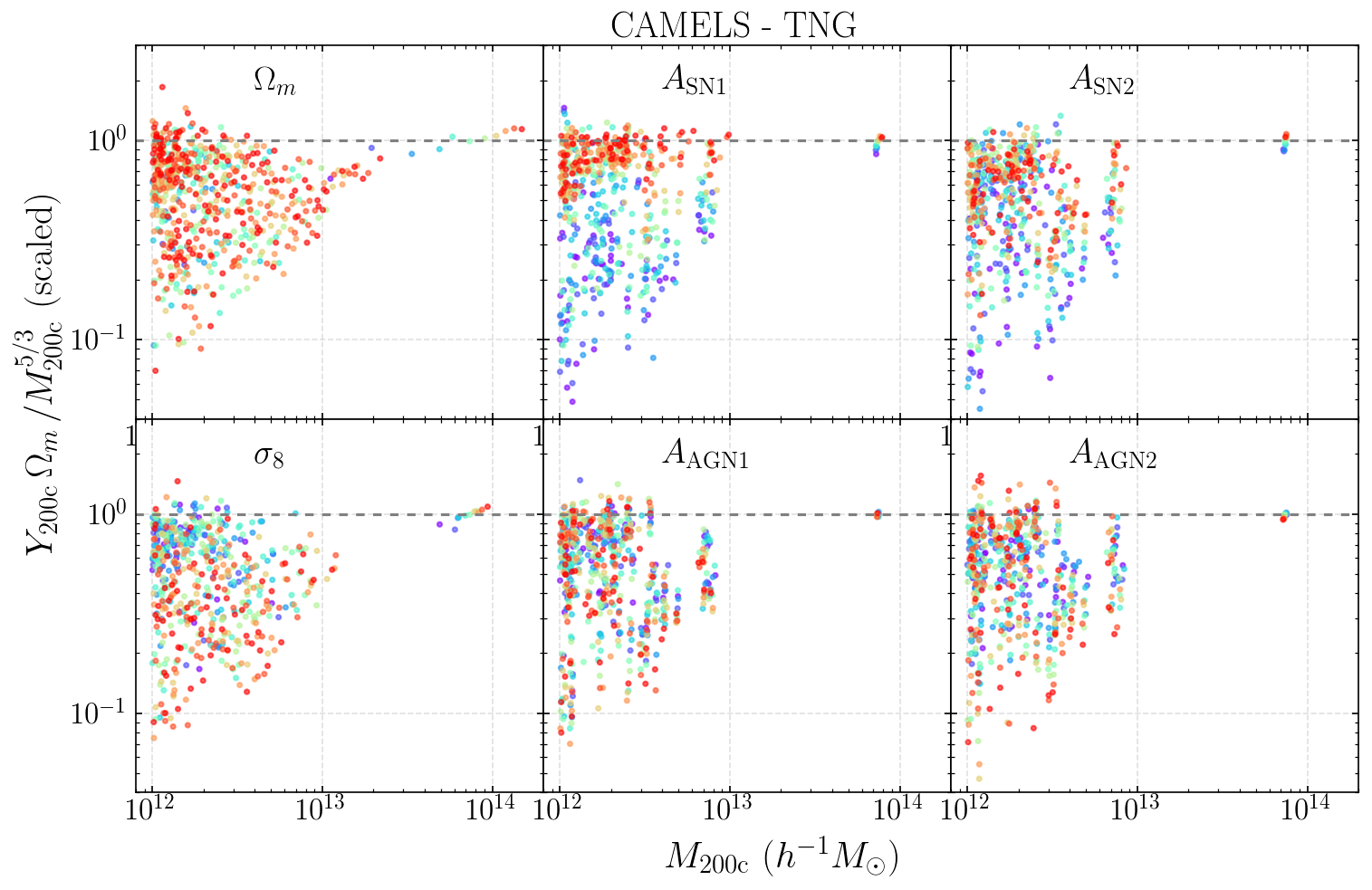}
\includegraphics[scale=0.25,keepaspectratio=true]{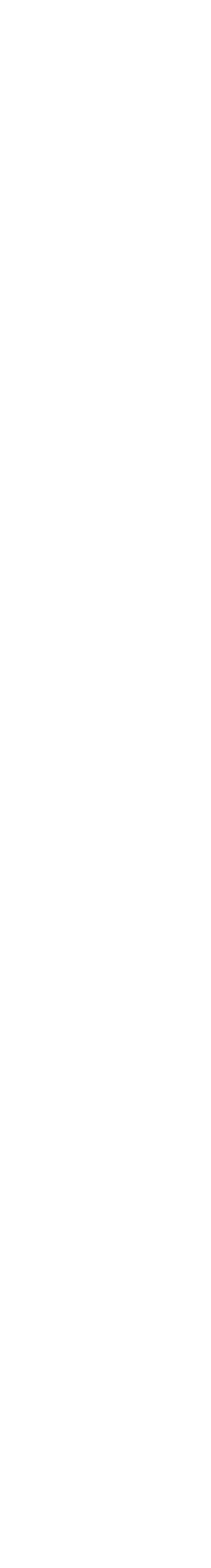}
\caption{$Y-M$ scaling relation for halos in the CAMELS-1P set at $z=0.27$. Each panel corresponds to halos from simulations with the same initial seeds but where only the parameter indicated in the upper-left is varied; the colors correspond to the parameter value. The power-law scaling relation approximately normalized to the high-mass clusters is shown by the dotted gray line. In most cases, we see that increasing the strength of AGN feedback moves the objects away from the virial theorem prediction, while the opposite is the case for SN feedback. A similar plot for the case when all the six parameters are simultaneously varied in a latin-hypercube fashion is given in Fig.~\ref{fig:AstroPropsLH}.}
\label{fig:AstroProps1P}
\end{figure*}

CAMELS contains two distinct simulation suites based on the code used to solve the hydrodynamic equations and implementation of the subgrid model: $(i)$ CAMELS-SIMBA, based on the GIZMO code \citep{Hop15,Hop17} employing the same sub-grid model as the flagship SIMBA simulation\footnote{SIMBA: \url{http://simba.roe.ac.uk/}} \citep{Dave2019};
$(ii)$ CAMELS-TNG, based on the AREPO code \citep{Spr10, Wei20} employing the same sub-grid model as the flagship IllustrisTNG simulations\footnote{IllustrisTNG: \url{https://www.tng-project.org/data/}} \citep{Nel19,Pil18,Spr18,Nel18,Nai18,Mar18,PilSprNel1801,Wei17,Vog14,Vog14b}.
Let us provide one example to highlight the substantial differences in these models: AGN feedback is implemented considering spherical symmetry in IllustrisTNG; while in SIMBA, it is modeled as collimated outflows and jets from AGN.
We denote the parameters in TNG as $A_i$ and those in SIMBA by $B_i$ \footnote{Note that the meaning of the feedback parameters differs substantially between IllustrisTNG and SIMBA and therefore we label the parameters differently.}.
$A_i=1$ ($B_i=1$) correspond to the fiducial values of the parameters used in the flagship TNG (SIMBA) models at the resolution considered in the CAMELS simulations.
The cosmological parameters are varied in a broad range: $\Omega_{\rm m}\in$[0.1 - 0.5], $\sigma_8\in$ [0.6 - 1.0], and the feedback parameters are varied between either $[0.25, 4.0]$ or $[0.5, 2.0]$.

For each of the two simulation codes, the CAMELS suite comprises the following sets of simulations which will be used in this work.
\begin{itemize}
\item LH (latin hypercube): 1,000 simulations in which cosmology and feedback parameters are varied on a latin hypercube, each run having a different random initial seed.

\item 1P (one parameter at a time): Simulations that only vary one parameter at-a-time and have the same initial random seed. This set contains 61 simulations.
\end{itemize}

CAMELS also has a third set called CV (from cosmic variance) where only the initial seeds were varied keeping the cosmological and astrophysical parameters fixed, but we have not used that in our analysis.

Because of the small volume of the CAMELS simulations (25 $\Mpc$ side), there is a dearth of high-mass objects in them. Therefore, we also use the
large boxes of the flagship TNG and SIMBA simulations.
We use a box with 300 Mpc side length from the TNG300-1 simulation (hereafter TNG300), and similarly, a box with 100 $\Mpc$ side from the SIMBA simulation (SIMBA100). Throughout our study, we use the cluster/group samples at $z=0.27$. We picked this redshift because it lies in the range of the DESI BGS forecasts ($0.2 \lesssim z \lesssim 0.3$) that we will later use in this paper.

Let us now discuss the cluster/group properties that we use. The procedure of calculating these properties from the simulation data is similar to our companion paper \citep{Wad22}.
For all the simulations, we choose the centers of halos to be the locations of the minimum gravitational potential within the FOF (friends of friends) volume.
 We use the boundary $R_\textup{200c}$ to define the cluster radii\footnote{\RT\ is the radius enclosing an overdensity $\Delta = 200$ with respect to the critical density of the Universe.}. $M_\textup{200c}$ is the mass of all the particles (dark matter, gas, stars and black holes) within $R_\textup{200c}$ of the halo center.
 
CMB photons are scattered by high energy electrons in the plasma inside clusters because of the inverse Compton scattering effect; this leads to a shift in the energy of CMB photons. Such a shift is directly measured in SZ surveys and is typically parameterized by the integrated Compton-$y$ parameter ($Y_\mathrm{SZ}$). In this paper, we consider a 3D analogue of $Y_\mathrm{SZ}$, given by,
\be
Y_{200c}= \frac{\sigma_\textup{T}}{m_e c^2}
\int_0^{R_{200c}} P_e (r)\, 4\pi r^2 dr
\label{eq:Y200}
\ee
where $\sigma_\textup{T}$ is the Thomson cross section, $m_e$ is the electron mass, $P_e (r)$ is the spherically-averaged electron pressure profile, and $c$ is the speed of light. We calculate the cluster ionized gas mass as
\be
M_\mathrm{gas}(r<R)= \frac{2}{1+X_H}\, m_p \int_0^{R} n_\textup{e} (r)\, 4\pi r^2 dr
\label{eq:Mgas}
\ee
where $n_e(r)$ is the spherically-averaged free electron number density profile, $X_H=0.76$ is the primordial neutral hydrogen fraction, and $m_p$ is the proton mass\footnote{We have derived $M_\mathrm{gas}$ using $n_e$ (instead of naively summing over masses of gas particles) in order to roughly mimic the $M_\mathrm{gas}$ measurements from X-ray surveys (where $n_\textup{e} (r)$ is derived by de-projecting of X-ray surface brightness profiles \citep{Voe04,Cro06}). Note however that we have ignored effects of gas clumping which can affect measurements of $M_\mathrm{gas}$ in X-ray surveys; a proper treatment of the gas clumping effect requires making mock X-ray maps (see e.g., \cite{Nag11, Ave14}) and is beyond the scope of this paper.}.
 
The stellar mass $(M_*)$ is calculated by summing over of the masses of all the star particles within \RT. Note that this quantity represents thus the total stellar mass in the cluster, not the stellar mass of the central galaxy. We have used the group\_particles code\footnote{\url{https://github.com/leanderthiele/group_particles}} to calculate the pressure, gas density and stellar density profiles from the simulation data.


\section{$Y-M$ scaling relation}
\label{sec:YM}

\begin{figure}
\centering
\includegraphics[scale=0.55,keepaspectratio=true]{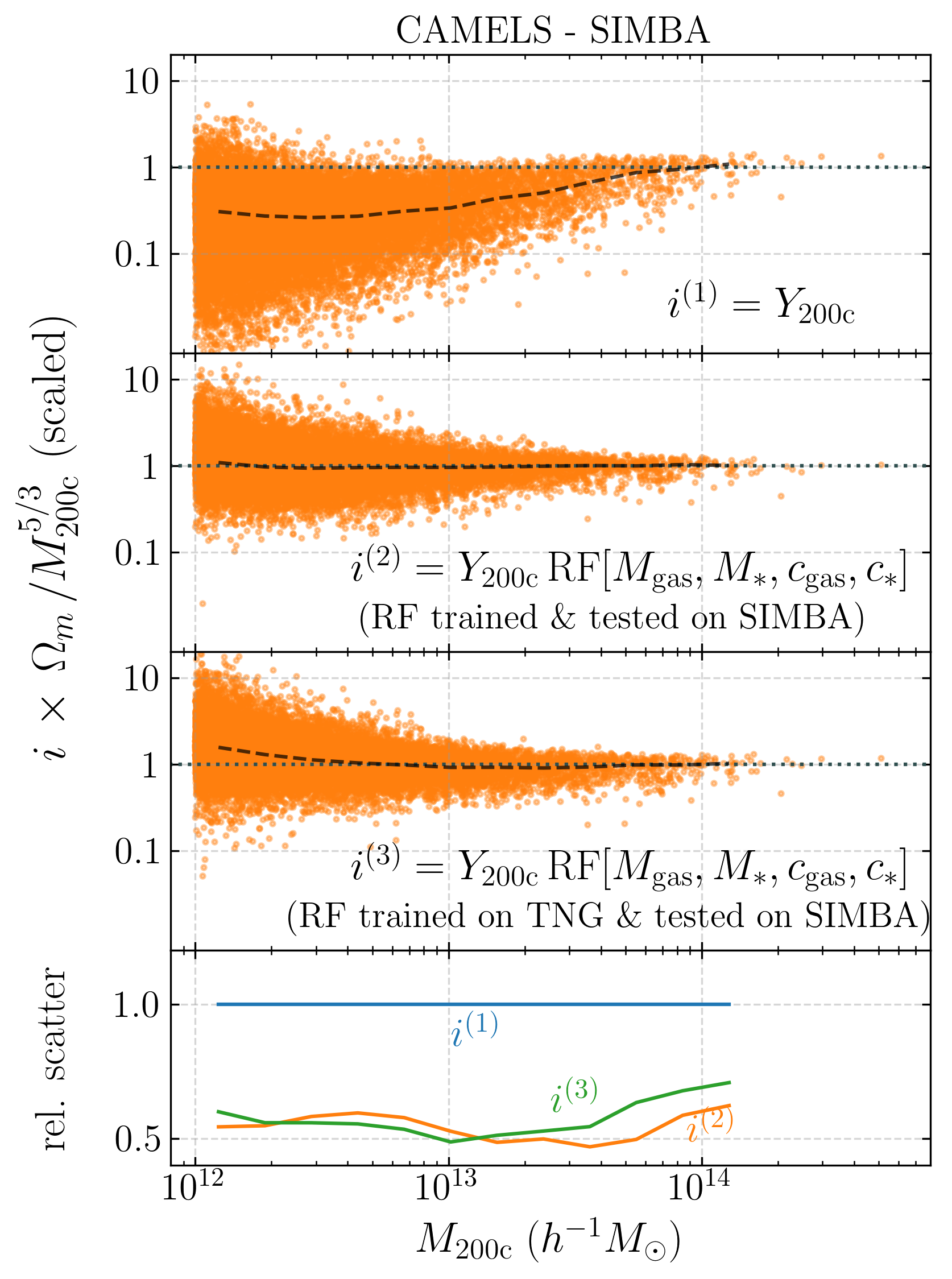}
\caption{A normalized version of the $Y-M$ relation from the latin-hypercube set of the CAMELS-SIMBA simulations is shown in the top panel. The middle two panels show the effect of adding a correction factor using the random forest regressor (RF). The third panel from top compares the generalization performance of the RF beyond its training set. The dashed black line shows the median of the relations, while the standard deviation is compared in the bottom panel. $M_*$ ($M_\mathrm{gas}$) is the stellar (gas) mass within \RT, $c_\mathrm{gas}$ corresponds to concentration of gas and is given by $M_\mathrm{gas}(r\, <\, R_{200c}/2)/M_\mathrm{gas}(r\, <\, R_{200c})$ (similarly, $c_*$ corresponds to the stellar mass concentration). A similar plot for CAMELS-TNG instead of CAMELS-SIMBA is in Fig.~\ref{fig:RF_TNG}. Overall, the RF augmented predictions have a substantially smaller deviation from a power-law relation and also have a smaller scatter.}
\label{fig:RF}
\end{figure}

\begin{figure*}
\centering
\includegraphics[scale=0.55,keepaspectratio=true]{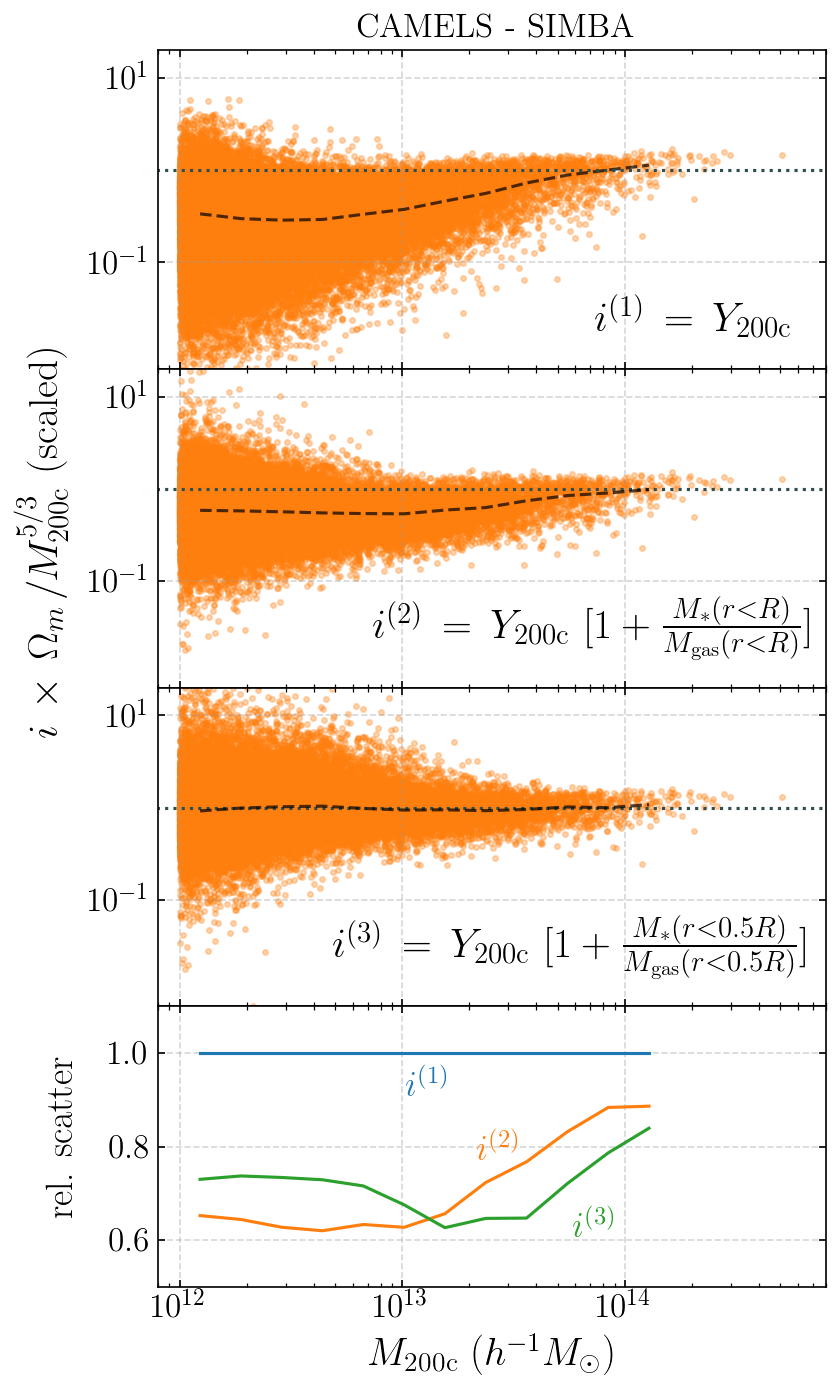}
\includegraphics[scale=0.55,keepaspectratio=true]{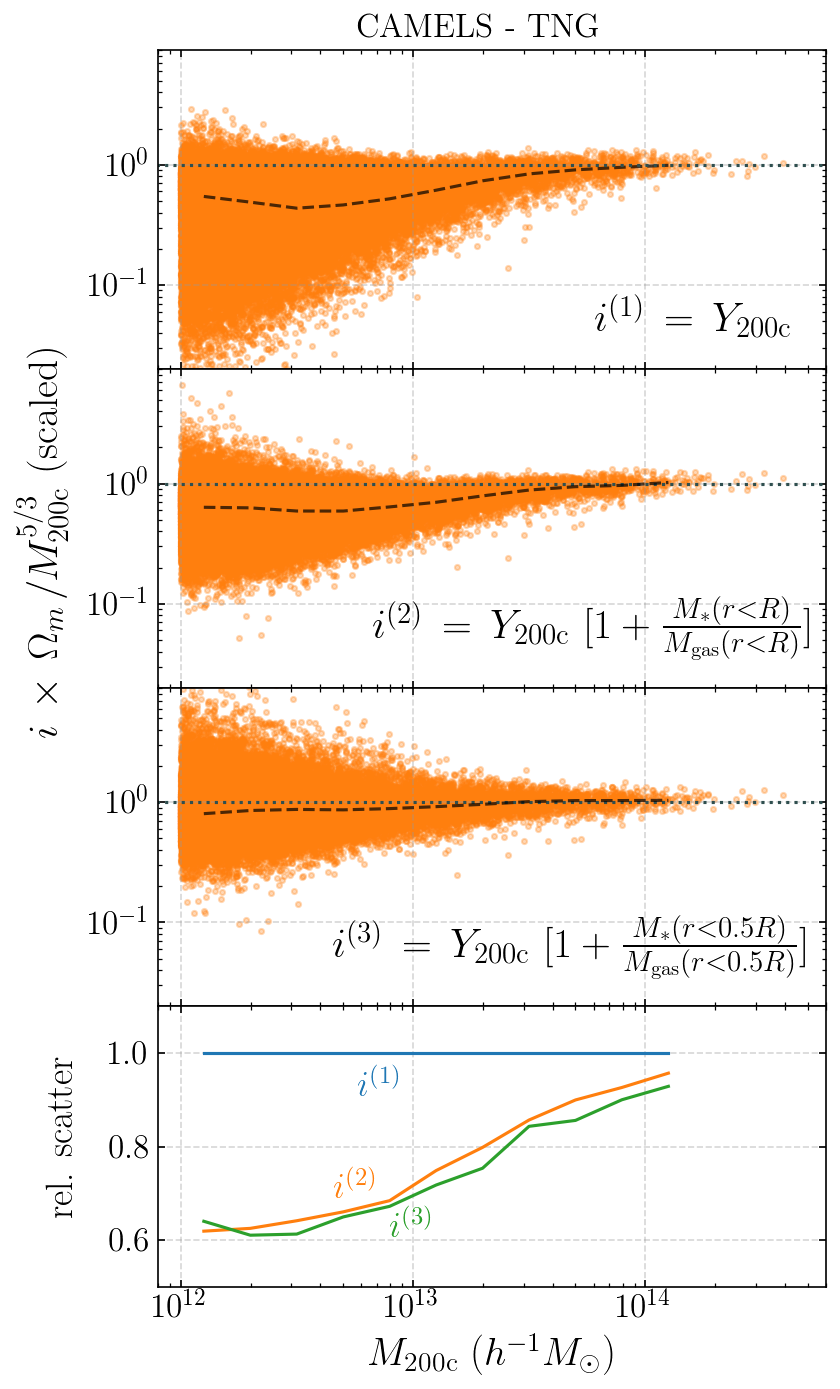}
\caption{Same as Fig.~\ref{fig:RF}, but comparing results from symbolic regression in the middle two panels (we have denoted \RT\ as $R$). Upon augmenting $Y_\textup{200c}$ with the ratio $M_*/M_\textup{gas}\, |_{r<R_{200c}/2}$, the relation remains close to a power law for much lower halo masses as compared to $Y-M$. Note that the improvements due to the augmented relations are robust to changing both the sub-grid model and feedback parameter strengths. Our relations can enable accurate mass estimation of low-mass clusters and galaxy groups using observational properties inferred from SZ, galaxy and X-ray surveys.}
\label{fig:Y_Mgal}
\end{figure*}

\begin{figure}
\centering
\includegraphics[scale=0.5,keepaspectratio=true]{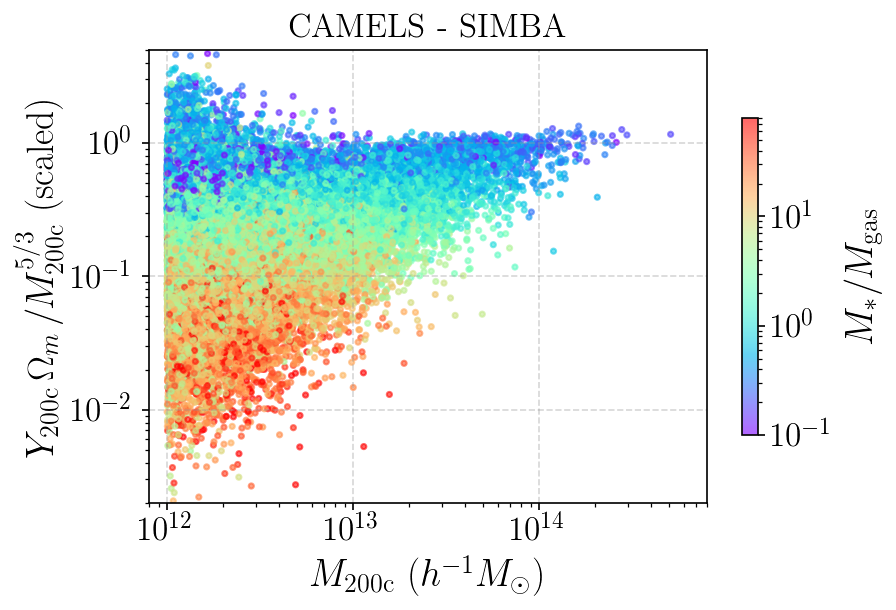}
\includegraphics[scale=0.5,keepaspectratio=true]{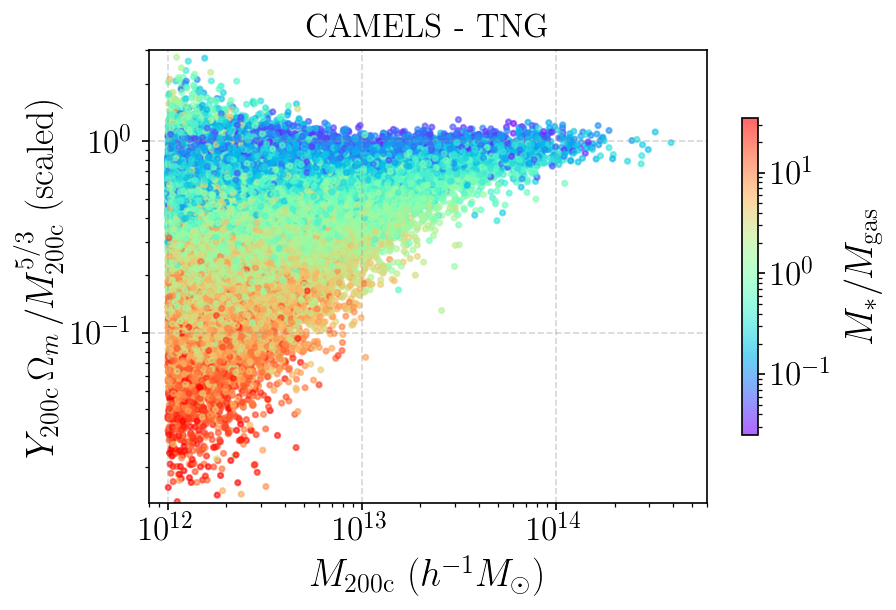}
\caption{As a cross-check to the results in Fig.~\ref{fig:Y_Mgal}, we plot the $Y-M$ relation as a function of $M_*/M_\textup{gas}\, |_{r<R_{200c}/2}$. The deviation from the self-similar relation indeed has a strikingly strong anticorrelation with $M_*/M_\mathrm{gas}$ for both TNG and SIMBA.}
\label{fig:StarGasRatio}
\end{figure}

Using the virial theorem, one can derive simple power-law scaling relations between various properties of clusters \citep{Kai86,KraBor12}.
For example, the scaling relation between cluster mass and temperature at a fixed redshift is given by $T \propto M^{2/3}$ (see Eq.~8 of \cite{Bry98}). Similarly, the scaling relation for the gas mass of a cluster is $M_\mathrm{gas} \propto \Omega_b/\Omega_m\, M$. Writing $Y \propto M_\mathrm{gas} T$, we can obtain

\be
\Omega_m Y= c_0\, M^{5/3}
\label{eq:YM}\ee
where $c_0$ is a constant whose value for a particular sample is calibrated using high-mass halos in that sample (the high-mass halos are used because the relation deviates from a power-law for low masses, as we will discuss in the next subsection). It is worth mentioning that the normalization constant is obtained phenomenologically (typically using CMB lensing or optical lensing measurements) rather than directly using predictions from hydrodynamic simulations.
Note that in the CAMELS simulations, $\Omega_b$ is fixed but $\Omega_m$ is allowed to vary. We have therefore absorbed the $\Omega_b$  dependence in the constant while showing the $\Omega_m$ dependence explicitly. There are also additional factors of redshift and $\Delta$ (the overdensity with respect to the critical density of the Universe), which we have ignored as we only consider halos at a fixed redshift in our analysis. 

\subsection{Dependence on astrophysical feedback}
\label{sec:Feedback_YM}

The virial theorem assumes that the only source of energy input into the intra-cluster medium is gravitational.
In order to study the deviation from the self-similar relation in Eq.~\ref{eq:YM} as a function of astrophysical feedback parameters, we use the CAMELS 1P set to analyze the ratio $(Y \Omega_m/c_0\, M^{5/3})$ in Fig.~\ref{fig:AstroProps1P}. We see that increasing the AGN feedback strength moves the objects away from the virial theorem prediction, while the opposite is the case for SN feedback.
We think that this happens because the jets/outflows from AGNs can efficiently eject the gas from halos, and hence we see a reduction in $Y$. 
The case is different for SN as SN driven outflows have lower velocities compared to those from AGNs, and thus SN winds are better confined by hydrodynamic drag and radiative losses. Therefore, SN outflows do not eject gas efficiently from halos, but rather heat the gas. Such heating prevents the reduction in $Y$ because of two reasons: $(i)$ gas cannot be efficiently converted to stars, $(ii)$ the growth of central black holes (which drive the AGN outflows) is impeded \citep{Boo13, Ang17}.

In Fig.~\ref{fig:AstroProps1P}, apart from the feedback parameters, $\Omega_m$ and $\sigma_8$ also seem to have an impact on the $Y-M$ relation. However, note that the cosmological parameters in CAMELS are varied in a very wide range ($\Omega_{m}\in$[0.1 - 0.5], $\sigma_8\in$ [0.6 - 1.0]), as compared to the errorbars from current surveys, e.g., Planck, which are at the percent level $(\Omega_m = 0.3147 \pm 0.0074,\ \sigma_8 = 0.8101 \pm 0.0061)$ \citep{Aga18}. The $Y-M$ relation is affected very weakly when cosmological parameters are varied within the Planck errorbars. Therefore measurement of $Y-M$ relation will not be a strong probe of cosmology, but does give strong constraints on baryonic feedback, as we will later study in section~\ref{sec:Results2}.

\section{Machine learning tools}
\label{sec:ML}

As discussed in Sec.~\ref{sec:intro}, our goal is to use machine learning tools to model the following function
\be
 f(\{i_\textup{obs}\}) = \frac{c_0\, M^{5/3}} {Y \Omega_m}~,
\label{eq:ML}
\ee
with the goal of finding a new relationship $Y\, f(\{i_\textup{obs}\}) - M$ that is robust to feedback effects.
We use a combination of two machine learning tools: a random forest regressor and symbolic regression.

Symbolic regression (SR) is a technique that approximates the relation between input and output variables through analytic mathematical formulae. It is worth mentioning that SR has been used in various astrophysical applications, e.g., \citep{Wad22,CraSan20,WadVil20b, Del21,Sha21,astroexample2,astroexample3,Ber21,VilAngGen20,Lem22,Won22, Sha23, Bar22, Bar23, Arj20}.

The advantage of using SR over other machine learning regression models 
is that it provides analytic expressions which can be readily generalized, and which facilitate the  interpretation of the underlying physics.
However, a major downside of SR is that the dimensionality of the input space needs to be relatively small. To overcome this, we use a similar method as that in \cite{Wad22, WadVil20b} to first use a random forest regressor (RF) to obtain an indication of the minimal set of parameters $\{i_h\}$ that performs fairly well (i.e., to use RF as a feature selector). Our motivation of using a decision-tree based tool like RF instead of a deep neural network is that decision trees are comparatively much faster to train, and they do not require access to GPUs. Furthermore, for datasets which are not extremely high dimensional, decision trees can achieve similar performance as neural networks (for a schematic comparison of different ML tools, see Fig.~1 of \cite{Wad22}).

 We use the symbolic regressor based on genetic programming implemented in the publicly available \textsc{PySR} package\footnote{\label{PySR}\textsc{PySR}: \url{https://github.com/MilesCranmer/PySR}} \citep{pysr}. For RF, we use the publicly available package \texttt{Scikit-Learn}\footnote{Random forest: \url{https://scikit-learn.org/stable/modules/generated/sklearn.ensemble.RandomForestRegressor.html}}.



\section{Results I: Reducing deviation from self-similarity}
\label{sec:Results1}

As discussed in section~\ref{sec:ML}, we train the RF in order to narrow down the parameter set for $\{i_\textup{obs}\}$ in Eq.~\ref{eq:ML} (the goal of the RF is to model the ratio $c_0\, M^{5/3}/ Y \Omega_m$ using different $\{i_\textup{obs}\}$ input parameter sets). We start by using the data of halos in the LH set of CAMELS-SIMBA. To avoid overfitting, we divide the data into a training set containing half of the halos to train the RF, and the rest of the data is used in testing the RF. We find that the RF gives the best results when $\{i_\textup{obs}\} = \{M_*$, $M_\mathrm{gas},c_*$, $c_\mathrm{gas}\}$ are provided as input ($c_\mathrm{*}$ and $c_\mathrm{gas}$ correspond to concentration of stars and gas, see the caption of Fig.~\ref{fig:RF}). We show the results for the CAMELS-SIMBA test set in Fig.~\ref{fig:RF}. We find that introducing additional parameters in the $\{i_\textup{obs}\}$ set (e.g., the richness of halos) provides little additional improvement in the RF performance (we show the importance of the different features for the RF prediction in Fig.~\ref{fig:FeatureImp}). We leave further details of the feature selection analysis to Appendix~\ref{apx:RF}. Apart from the mean of the different relations, we also quantify their scatter in the bottom panel of Figure~\ref{fig:RF} (the scatter is given by the standard deviation of the log of the ratio shown in the y-axis of the top panels). We do not compare the scatter for the very high-mass end ($M\gtrsim 10^{14} \Ms$) as there are very few clusters available to calculate the scatter robustly.

In order to test how well the RF can generalize beyond its training set, we show in the third panel of Fig.~\ref{fig:RF} the case when the RF is trained on groups from the TNG simulations and tested on groups in the SIMBA simulations. It is interesting to see that the modelling by RF is fairly robust to change in the sub-grid prescriptions.
A similar plot to Fig.~\ref{fig:RF} but corresponding to testing the RF on CAMELS-TNG data instead of CAMELS-SIMBA is in Fig.~\ref{fig:RF_TNG}. 

Next, we try to figure out if we can get a comparable performance with a simple function taking some/all of these properties as inputs. Upon manual trial and error, we find the relation
\be
M^{5/3} \propto Y \Omega_m\bigg( 1+ \frac{M_* (r<R_{200c})}{M_\textup{gas}(r<R_{200c})} \bigg)
\label{eq:YM2}
\ee
gives results closer to a power-law as compared to the traditional $Y-M$ for both TNG and SIMBA (see figure~\ref{fig:Y_Mgal}). Roughly, we see by the following substitution
\begin{equation*}
\begin{split}
Y\, \left(1+\frac{M_*}{M_\mathrm{gas}}\right) \propto&\ M_\mathrm{gas} T_\mathrm{gas}\, \left(1+\frac{M_*}{M_\mathrm{gas}}\right)\\ \propto&\ (M_*+M_\mathrm{gas})T_\mathrm{gas}
\end{split}\end{equation*}
that our new relation depends on the sum of the gas and stellar mass. It is therefore robust to feedback processes governing the conversion of gas to stars (note, however, that it is not robust to feedback corresponding to ejection of gas from the halos).

We then use symbolic regression (SR) on the compressed $\{i_\textup{obs}\} = \{M_*$, $M_\mathrm{gas},c_*$, $c_\mathrm{gas}\}$ set as input to see if we can obtain an even better expression. Upon initial runs of SR, we found the parameter combination $M_\mathrm{gas} c_\mathrm{gas}$ (which essentially equates to $M_\textup{gas}(r<R_{200c}/2)$) frequently appears in the equations output by SR. The best expression we found from SR reduces to
\be
M^{5/3} \propto Y \Omega_m\bigg( 1+ \frac{M_* (r<R_{200c}/2)}{M_\textup{gas}(r<R_{200c}/2)} \bigg)\, ,
\label{eq:YM3}
\ee
and is much closer to a power-law relation as compared to the traditional $Y-M$ and also to Eq.~\ref{eq:YM2} (see figure~\ref{fig:Y_Mgal}).
A rough explanation for this could be that baryonic feedback affects the inner part of the halo more than the outskirts, and therefore the term in Eq.~\ref{eq:YM3} provides a better ``correction'' to the feedback effects as compared to Eq.~\ref{eq:YM2}. It is important to note that the our new mass proxy not only has lower bias, but also significantly lower scatter at the low-mass end, as seen from the bottom panel of figure~\ref{fig:Y_Mgal}.

To explicitly see how the variation in $Y/M^{5/3}$ is correlated with \Msg, we show Figure~\ref{fig:StarGasRatio} corresponding to the CAMELS LH set. It is worth mentioning that a recent study by \cite{Yan22} reported that the hot gas content of halos was the primary driver of the break in the $Y-M$ relation. In our case, we find that augmenting the $Y-M$ relation with the ratio $M_*/M_\mathrm{gas}$ performs better as compared to using $M_\mathrm{gas}$ alone. For the case of groups and clusters of galaxies, $M_\mathrm{gas}$ can be measured in X-ray surveys (see e.g., \cite{Sun09}) and $M_*$ from galaxy surveys (see e.g., \cite{Bla05, Pal20}). Our new mass proxy in Eq.~\ref{eq:YM3} can therefore exploit the multi-wavelength data of these objects to accurately infer their halo mass.

Note that we find additional equations from SR which have a better performance than Eq.~\ref{eq:YM3}, but are more complex; some of these equations as shown in Fig.~\ref{fig:Y_Mgal2} of the Appendix. 
Our aim was to include in Fig.~\ref{fig:Y_Mgal} the simplest expressions which have a relatively good performance. We mention some possible ways to improve the SR performance in Appendix~\ref{apx:SR}.

Fig.~\ref{fig:Y_Mgal} was made for the LH set (where both the cosmology and feedback parameters were varied simultaneously). We also show an alternate version of the figure for the 1P set in Fig.~\ref{fig:Y_Mgal_1P}, where we only varied the feedback parameters and fixed the cosmological parameters to their fiducial values ($\Omega_m=0.3, \sigma_8=0.8$). We again find a similar level of improvement when using our new expressions.

%


\begin{figure*}
\centering
\includegraphics[scale=0.58,keepaspectratio=true]{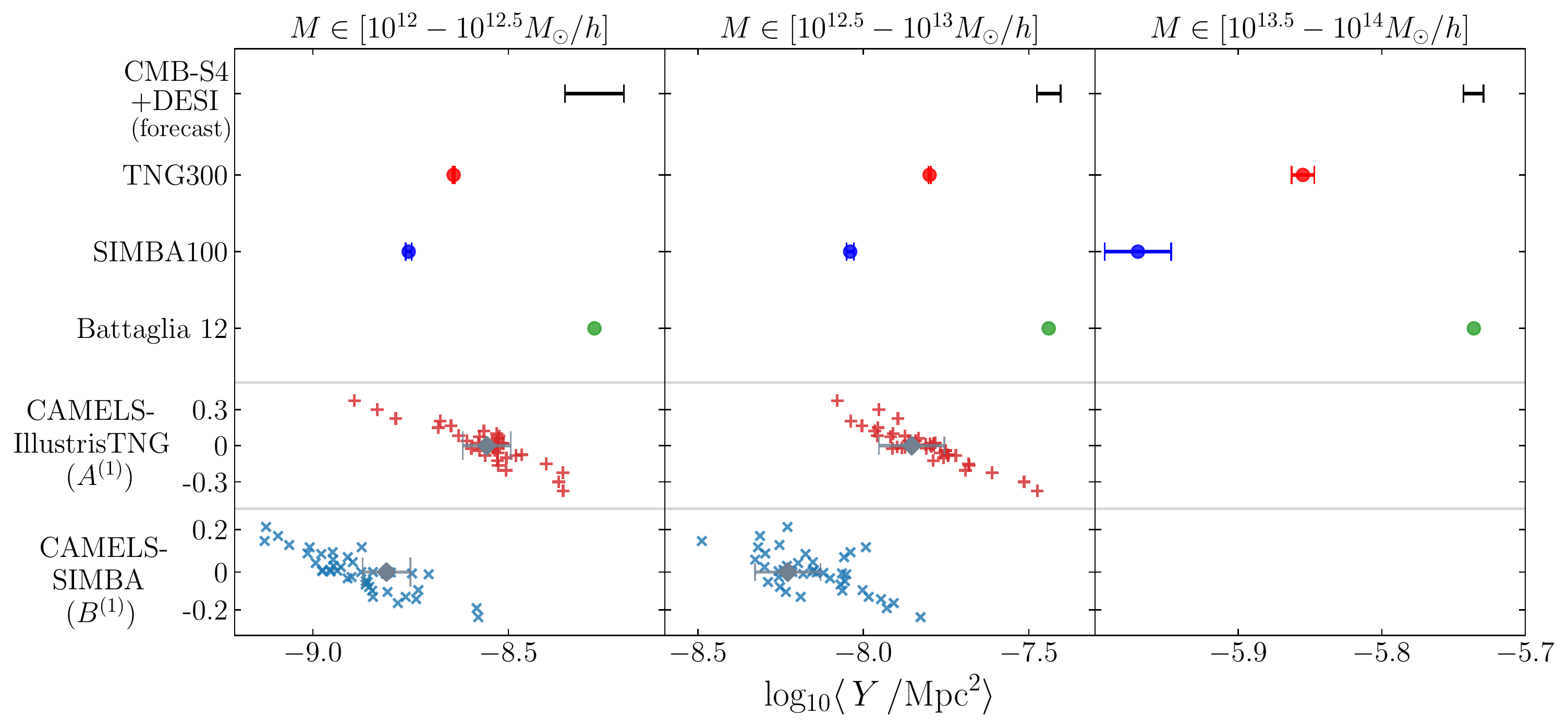}
\caption{Measurement of mean of $Y$ of halos in three different mass bins from the simulation boxes of IllustrisTNG300, SIMBA100 and CAMELS 1P-set, and predictions based on the pressure profile model in \citet{BatBon12b}.
Note that all the CAMELS simulations shown in this figure have the same cosmology and initial seeds, and therefore enable us to explicitly compare the variations due to different feedback prescriptions (the y-coordinates of CAMELS simulations correspond to log$_{10} (A^{(1)}_\mathrm{IllustrisTNG})$ and log$_{10} (B^{(1)}_\mathrm{SIMBA})$ in Eq.~\ref{eq:ParamConstraints}). For each simulation case, we also show the cosmic variance errorbars corresponding to the finite volume of the simulations.
We show forecasts from \citet{Pan20} for a CMB-S4-like SZ survey and a DESI-like galaxy survey. Simultaneous measurements of $Y$ and $M$ from upcoming surveys could therefore rule out a major part of the parameter space corresponding to supernova \& AGN feedback models used in current state-of-the-art hydrodynamic simulations.}
\label{fig:mean_Y}
\end{figure*}

\begin{figure*}
\centering
\includegraphics[scale=0.26,keepaspectratio=true]{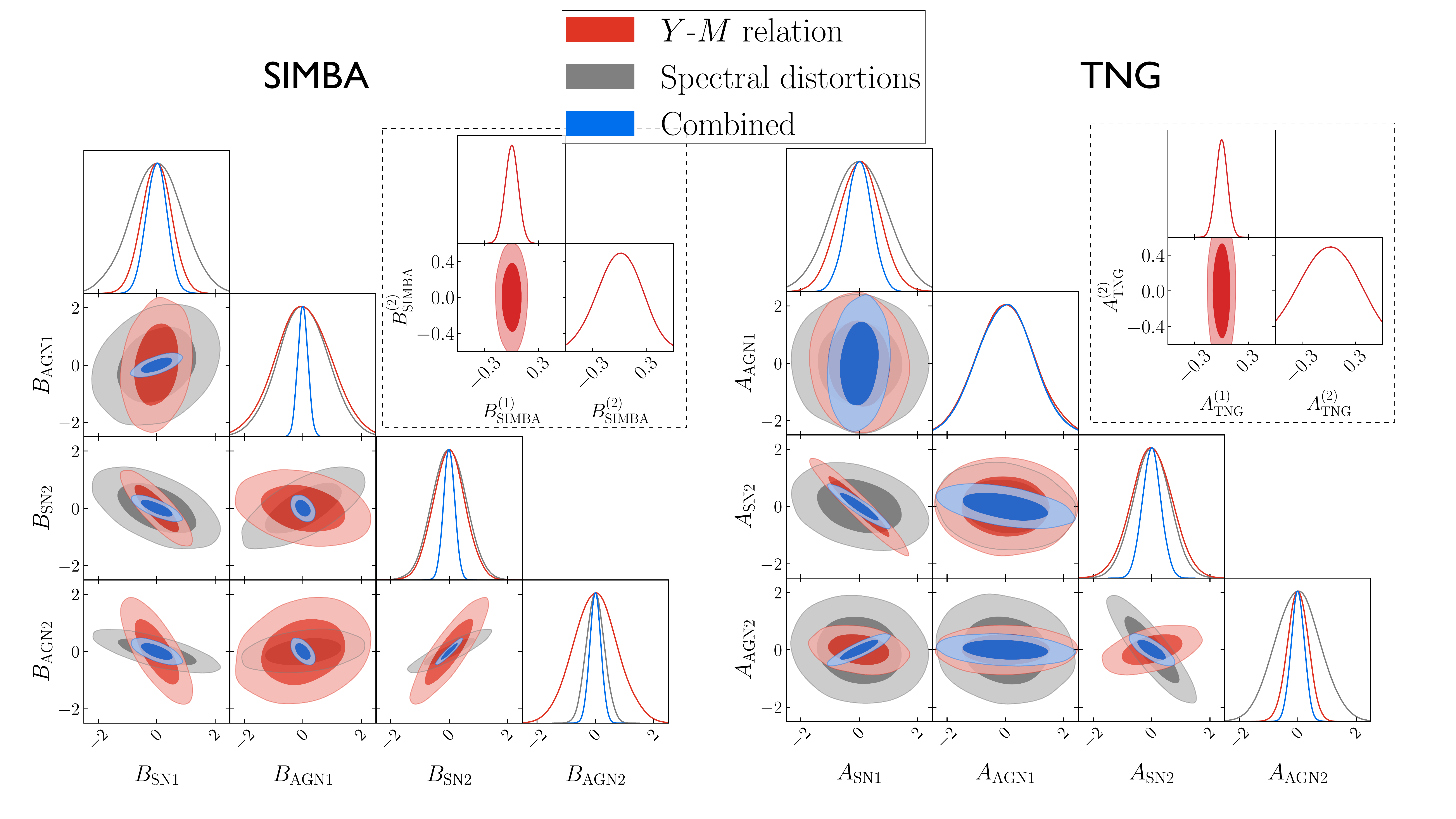}
\caption{Constraints on feedback parameters using $Y-M$ measurement forecasts from \citet{Pan20} for a CMB-S4-like CMB survey and a DESI-like galaxy survey. We have performed the analysis for the two CAMELS suites (SIMBA and TNG) separately. We show log$_{10}$ of the parameter values on all the axes and have used weak Gaussian priors ($1\sigma=1$) for each of the parameters (see the text for details).
Although constraints on some parts of the parameter space are dominated by the priors, there are particular parameter combinations which are strongly constrained by $Y-M$ measurements; we show these combinations in the inset plots (see  Eq.~\ref{eq:ParamConstraints} for their explicit form). We also show the combination with the forecasts of \citet{Thi22} for a PIXIE-like survey of CMB spectral distortions.}
\label{fig:triangle}
\end{figure*}

\begin{figure*}
\centering
\includegraphics[scale=0.5,keepaspectratio=true]{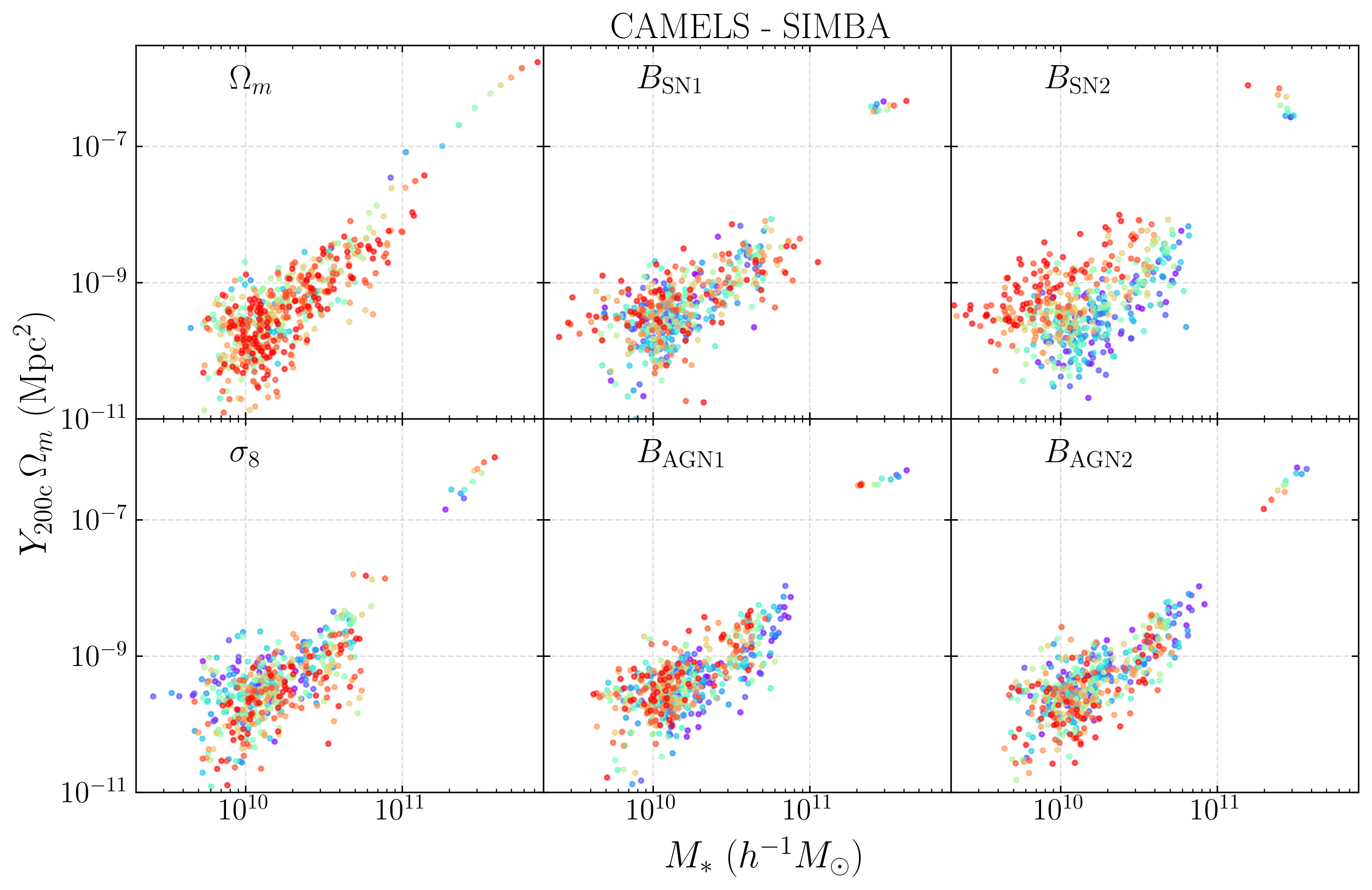}
\includegraphics[scale=0.25,keepaspectratio=true]{Figs/colorbar.pdf}
\includegraphics[scale=0.5,keepaspectratio=true]{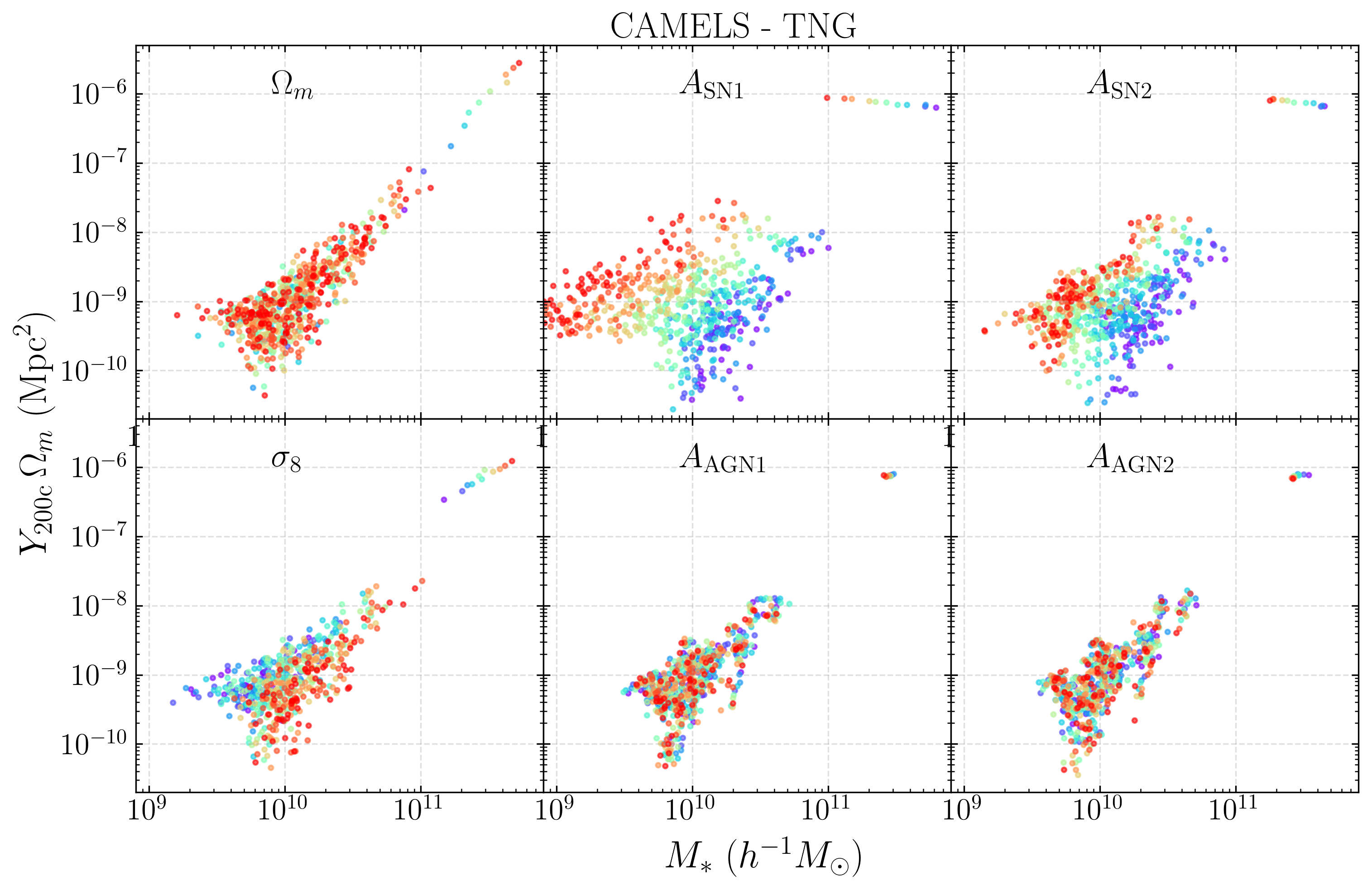}
\includegraphics[scale=0.25,keepaspectratio=true]{Figs/blank.pdf}
\caption{Same as Figure~\ref{fig:AstroProps1P}, but instead showing the relation of $Y$ with stellar mass ($M_*$). The advantage of using $M_*$ is that it is, in some cases, easier to infer from galaxy data as compared to $M_{200c}$. We see that most of the supernova feedback parameters have a strong impact on normalization of the $Y-M_*$ relation (while AGN feedback has either a small effect or roughly moves objects along the power-law slope). As this behavior is different from $Y-M$ (for which the SN and AGN have roughly degenerate effects as both affect its normalization, see Fig.~\ref{fig:AstroProps1P}), measuring $Y-M_*$ can provide complementary information about baryonic feedback.} 
\label{fig:Y-M*}
\end{figure*}

\section{Results II: constraining baryonic feedback}
\label{sec:Results2}

\subsection{Constraints on sub-grid prescriptions}

We have shown in Fig.~\ref{fig:AstroProps1P} that the baryonic parameters have a strong effect on the magnitude of deviation from the self-similarity in the $Y-M$ relation. This deviation could be measured in current and upcoming surveys by simultaneously measuring both $Y$ and $M$ of halos. We explore how such measurements can help in distinguishing different sub-grid prescriptions (e.g., TNG and SIMBA), and also in determining strength of AGN/SN feedback parameters.

The mean of the $Y-M$ relation can be measured by cross-correlating tSZ maps with either weak lensing maps or galaxy fields \citep{Hil18,Pan20,Gat21,Pan21, Osa18, Osa20}.
We use forecasts of $Y-M$ from cross-correlations from CMB-S4 and a DESI-like survey from \cite{Pan20} (hereafter \citetalias{Pan21}) to constraint baryonic feedback parameters in the CAMELS simulations.
We now briefly describe the ingredients used in the \citetalias{Pan21} forecast and refer the reader to their paper for further details. 
They used the \citet{Tin08} halo mass function and considered the redshift bin $z\in[0.2-0.3]$ and halo masses larger than $10^{12}\Ms$ for their analysis (these choices were made in order to very roughly reflect the mass and redshift coverage of the Bright-Galaxy-Sample (BGS) of DESI survey). The $Y$ maps are made assuming noise curves for CMB-S4. \citetalias{Pan21} reports two versions of the constraint on the $Y$-$M$ relation: one by identifying halos directly from a galaxy survey, and the other by measuring the galaxy-$y$ correlation and then relating the halos to galaxies assuming a standard HOD model. We use the forecasts from the former analysis, which are given in Table~\ref{tab:MeanY}. The mean value of the forecasts are based on the pressure profile model fitted to hydrodynamical simulations in \cite{BatBon12b} (which roughly corresponds to a power law $Y$-$M$ relation with slope $\sim$ -1.72). Note that the $Y-M$ constraints reported in \citetalias{Pan21} were originally derived for $R_{500c}$. We use their electron pressure profile model to rederive the $Y-M$ constraints for $R_{200c}$. As described in \citetalias{Pan21}, the covariance in the halo-y correlations is a sum of Gaussian and non-Gaussian components. The Gaussian component contributes mainly across all scales and is dominated by shot noise originating from the finite halo density and instrumental noise in the tSZ maps. Therefore, changing the fiducial pressure profile would have a sub-dominant impact on the inferred error in the $Y-M$ relation shown in Table~\ref{tab:MeanY}.



\begin{table}
\centering
\begin{tabular}{|p{2.5cm}||p{3cm}|}
\hline
Log$_{10}(M/M_\odot)$ & Log$_{10}(Y^{+1 \sigma}_{-1 \sigma})$ \\ \hline
$12.0-12.5$  & $-8.28^{+0.08}_{-0.07}$\\
$12.5-13.0$ & $-7.44^{+0.04}_{-0.03}$\\
$13.5-14.0$ & $-5.736^{+0.008}_{-0.006}$\\
\hline
\end{tabular}
\caption{Forecasts from \citet{Pan20} on the mean of the $Y-M$ relation using the the halo-$y$ cross-correlation corresponding to a CMB-S4-like SZ survey and a DESI-like galaxy survey. All the $Y$ values are in units of Mpc$^{2}$.}
\label{tab:MeanY}
\end{table}

We compare the results for the mean of the $Y-M$ relation from different hydrodynamic simulations in Fig.~\ref{fig:mean_Y}. Note that we calculate the mean $Y$ by averaging over the contribution from all halos in the indicated mass bin (we have not shown the variation among individual halos, which is relatively much larger). The CAMELS-1P simulations shown in the figure are for the same cosmology and initial seeds. Therefore, the different positions of the CAMELS points are solely due to changing baryonic feedback. The gray diamonds correspond to the CAMELS simulations which have parameters that are closest to the fiducial values used in the flagship TNG and SIMBA simulations ($A_i = B_i = 1$).
See \cite{Yan22} for a more detailed discussion of the physical effects underlying the differences between the $Y-M$ relation for TNG300 and SIMBA100 simulations.
We could not directly include the results in Fig.~\ref{fig:mean_Y} from CosmoOWLS simulations (Fig.~1 of \cite{LeB15}) as they been reported in the literature for $R_{500c}$ instead of $R_{200c}$.

We do not show results from CAMELS for the high halo mass bin ($M\in[10^{13.5}-10^{14}]\Ms$) because objects with these masses are found rarely in the small CAMELS boxes with side $25 \Mpc$. Therefore the results for this mass bin could be affected by cosmic variance. However, for low-mass bins, we have enough number of objects to compare the different simulations (for reference, the number of halos in each of the 1P set simulations for $M\in[10^{12}-10^{12.5}]\Ms$ ($M\in[10^{12.5}-10^{13}]\Ms$) bin is 32 (10)). As an additional check, we show the cosmic variance errorbar corresponding to the volume of the CAMELS simulations in Fig.~\ref{fig:mean_Y} and see that the CAMELS results are fairly consistent with the flagship TNG300 and SIMBA100 simulations for the case of fiducial feedback parameters ($A_i = B_i = 1$).


\subsection{Constraints on the strength of AGN/SN feedback}

To quantify the constraining power of the $Y-M$ relation on individual feedback parameters, we assume that the parameter likelihood is Gaussian. We perform a Fisher forecast for constraints using the $Y-M$ measurements in the two low-mass bins in our analysis: $10^{12}-10^{12.5}\Ms$ and $10^{12.5}-10^{13}\Ms$.
We calculate the derivatives of the logarithm of mean $Y$ for halos in the mass bin $i$, $\frac{\partial \log \overline{Y}_i}{\partial \log p_a}$, using the CAMELS 1P set, where $p_a$ correspond to the four feedback parameters used in the simulations (the cosmological parameters are fixed to their fiducial values ($\Omega_m=0.3, \sigma_8=0.8$) while calculating the derivatives).

 We numerically calculate the derivative using fourth-order interpolation (we find that our results have a weak dependence to the order of interpolation used, simulation suites with a finer variation in $p_a$ parameters will therefore be helpful for calculating the derivatives more robustly). The Fisher matrix corresponding to the $Y-M$ relation is given by \citep{Teg9711}:
\be
F_{ab}=\sum_{ij}^\mathrm{max} \frac{\partial \log \overline{Y}_i}{\partial \log p_a}\, \textup{Cov}^{-1}_{ij}\, \frac{\partial \log \overline{Y}_j}{\partial \log p_b}\,, \label{eq:Fisher}\ee
where Cov is the covariance matrix, \{$i,j$\} corresponding to mass bins, and \{$a,b$\} are indices corresponding to the four astrophysical parameters. We infer the diagonal elements of the covariance matrix from the errorbars given in Table~\ref{tab:MeanY} ($C_{ii} = [\Delta \log Y_i]^2$) and neglect the cross-correlation between the two mass bins\footnote{Note that we have used different values of $\Delta \log Y (= \Delta Y/Y)$ for TNG and SIMBA. This is because the mean $Y$ values for the simulations are different as seen in Fig.~\ref{fig:mean_Y}, and we conservatively use the same $\Delta Y$ for the two cases (e.g., 1$\sigma$ error in $\log_{10} Y$ is $\sim$0.22 (0.17) for SIMBA (TNG) for $M\in[10^{12}-10^{12.5}]\Ms$).}.
As we are trying to constrain four parameters but only have two data points, we have added a weak Gaussian prior ($\sigma_{\log_{10}p} = 1$) to make the Fisher matrix invertible. We perform our analysis separately for TNG and SIMBA and show the triangle plot in Fig.~\ref{fig:triangle}. As we have only two data points, we expect only two combinations of the feedback parameters to be constrained; we diagonalize the Fisher matrix and find these combinations to be:

\begin{equation}
\begin{split}
B^{(1)}_\text{SIMBA}& = B_\text{SN1}^{-0.06} B_\text{AGN1}^{-0.21}
                        B_\text{SN2}^{-0.81} B_\text{AGN2}^{+0.54}\,, \\
B^{(2)}_\text{SIMBA}& = B_\text{SN1}^{+0.89} B_\text{AGN1}^{-0.13}
                        B_\text{SN2}^{+0.22} B_\text{AGN2}^{+0.38}\,.
                        \\
A^{(1)}_\text{IllustrisTNG}& = A_\text{SN1}^{-0.63} A_\text{AGN1}^{-0.04}
                               A_\text{SN2}^{-0.69} A_\text{AGN2}^{+0.34}\,, \\
A^{(2)}_\text{IllustrisTNG}& = A_\text{SN1}^{+0.34} A_\text{AGN1}^{+0.02}
                               A_\text{SN2}^{+0.14} A_\text{AGN2}^{+0.93}\,,
\end{split}\label{eq:ParamConstraints}
\end{equation}

The corresponding 1$\sigma$ errorbars are:
\begin{align}
\sigma(B^{(1)}_\text{SIMBA})& = 3.2\%\,, &\sigma(A^{(1)}_\text{IllustrisTNG})& = 2.9\%\,, \nonumber\\
\sigma(B^{(2)}_\text{SIMBA})& = 11.3\%\,,   &\sigma(A^{(2)}_\text{IllustrisTNG})& = 16.2\%\,. \nonumber
\end{align}

Note that we have ignored the derivatives of $\bar{Y}$ with respect to cosmological parameters in our Fisher forecast. We have checked that if we include the derivatives with $\Omega_m$ \& $\sigma_8$, and also include as priors the Planck errorbars on them, our constraints on baryonic feedback parameters are affected negligibly.

Recent studies by \cite{Thi22} and \cite{Mos22} also forecasted the constraints on baryonic feedback parameters in CAMELS using CMB spectral distortions and kSZ/tSZ effect respectively. \cite{Thi22} showed that an experiment similar to PIXIE can impose percent level constraints on some combinations of feedback parameters. We show in Fig.~\ref{fig:triangle} their constraints and their combination with our analysis. We see that in some cases degeneracies are broken and the combined constraints are much tighter. 

Apart from $Y-M$, we explored alternative relations that could also be used to constrain feedback. Interestingly, the relation between $Y$ and $M_*$ is also sensitive to feedback strength, as seen in Fig.~\ref{fig:Y-M*}. Furthermore, we see that AGN and SN feedback affect the $Y-M_*$ relation differently: AGN feedback, roughly, tends to move objects along the power law, while SN feedback can substantially change the normalization of the relation. Inferring $M_*$ from optical galaxy surveys is easier than inferring $M_{200c}$ as one does not need to make assumptions about the galaxy halo connection (there is however an uncertainty in estimating of $M_*$ from galaxy spectra/photometry, but it is comparatively smaller, see the discussion in section~\ref{sec:future}). $Y-M_*$ has also been constrained in previous studies (see e.g., Fig.~4 of \cite{Ade13} for the case of locally brightest galaxies). It would be interesting to make a figure similar to Fig.~\ref{fig:mean_Y} but instead for $Y-M_*$ in order to gauge its constraining power, but we leave this to a future work.



\section{Discussion and conclusions}
\label{sec:Conclusions}

Upcoming arc-minute resolution CMB surveys like ACT, SO, CMB-S4 will be able to probe the integrated tSZ flux of low-mass clusters and galaxy groups with high accuracy. The SZ flux-mass relation ($Y-M$) for these objects deviates from the power-law prediction of the virial theorem. This is because the potential wells of low-mass halos ($M \lesssim 10^{14}\Ms$) are shallower, which makes the ionized gas more sensitive to feedback from active galactic nuclei and supernovae (Fig.~\ref{fig:AstroProps1P}).

We used a combination of random forest and symbolic regression to reduce this deviation in the $Y-M$ relation (Fig.~\ref{fig:RF}). We find a simple alternative relation: $Y(1+M_*/M_\mathrm{gas})-M$ has significantly better performance (Fig.~\ref{fig:Y_Mgal}): not only is our new relation close to a power-law to much lower halo masses, it also has significantly lower scatter than $Y-M$. Using the CAMELS suite of simulations, we tested that our new relation is robust against variations in not only baryonic feedback prescriptions but also cosmology and cosmic variance. 

In the 2020s, not just CMB surveys, but also galaxy surveys like DESI \citep{DESI}, Rubin \citep{Rubin}; and X-ray surveys like eROSITA \citep{Liu22,Chi22} will provide a wealth of multiwavelength data on galaxy groups and clusters.
Our new relation can enable accurate mass inference of low-mass cluster and groups of galaxies as it uses observable properties which are typically measured in these surveys ($M_*$ in galaxy surveys and $M_\mathrm{gas}$ in X-ray surveys). Our methodology of using machine learning tools can be useful for improving scaling relations in other areas of astrophysics where the relations either have a large scatter or deviate from a power law for particular systems. A couple of notable examples in this regard are: ($i$) the stellar to halo mass relation, which shows a break in the power law towards low-mass galaxies, see e.g., \cite{Wec18}; ($ii$) the scaling relation between black hole mass and bulge mass/velocity dispersion, which shows an increased scatter at low black hole masses, see e.g., \cite{Gre20}).
ML tools could be useful to improve the domain of validity of such relations.

In the second part of the paper, we used measurements of $Y-M$ relation to constrain the feedback prescriptions. One of the ways of constraining the $Y-M$ relation is by cross-correlating thermal SZ maps with galaxy maps. 
We showed that forecasts of such measurements from upcoming surveys like CMB-S4 and DESI can be used to place percent-level constraints on certain combinations of feedback parameters used in simulations (Fig.~\ref{fig:triangle}); they also have the potential to discriminate between different sub-grid models, e.g., between IllustrisTNG and SIMBA (Fig.~\ref{fig:mean_Y}).
Finally, we explored relations other than $Y-M$ and found that $Y-M_*$ can provide complementary constraints on feedback as compared to those from $Y-M$ (Fig.~\ref{fig:Y-M*}). Another advantage of $Y-M_*$ is that, in some cases, it is easier to observationally constrain than $Y-M$.

\subsection{Future work:}
\label{sec:future}

 
It is worth noting that we have used $M_*$ (the total stellar mass within $R_{200c}$) of objects from the simulation data in multiple results in this paper. In practice, however, $M_*$ is not directly observable but has to be inferred from spectra or photometry of galaxies in the halo. This can involve assumptions about redshift, star formation history, the initial mass function, the dust content, and stellar evolution models. However, the overall uncertainty in inferring $M_*$ is not very large, as one can see from the following estimates given in recent studies.
Using DES photometry alone, \cite{Pal20} estimate that the errors on $M_*$ are $\sim 0.2-0.3$ dex for $0.2<z<0.6$ (see their Fig.~3). Combining spectroscopic information along with photometry however improves the predictions, e.g., \cite{Hah22} estimate $\sim 0.1$ dex error bar on $M_*$ for the upcoming DESI BGS survey (it is worth adding that the estimated $M_*$ error in the locally brightest galaxies sample of SDSS has also been $\sim$0.1 dex \citep{Bla05}). Hydrodynamic simulations provide mock multi-band photometry data of simulated galaxies. One could therefore directly use galaxy photometric magnitudes instead of $M_*$ as inputs to machine learning models, but we leave this for a future study.
 
 We have used three dimensional cluster properties (e.g., \YT, or $M_\mathrm{gas}$ within \RT) in this paper. In reality, however, only projected properties (e.g., $Y_\mathrm{cylindrical}$) can be measured. We leave translating our results to projected parameters to a future study. It is worth noting that a recent study by \cite{Yan22} finds that effects of projection and finite beam size of the survey can degrade the sensitivity of $Y-M$ relation to feedback, however this degradation effect is weak for the case of upcoming arc-second resolution CMB surveys.

We have only considered groups/clusters at a particular redshift $z=0.27$ in our analysis. For the results in Fig.~\ref{fig:Y_Mgal}, we have checked that the radius at which the \Msg\ ratio gives optimal results can differ from $R_{200c}/2$ for different redshifts. However, we find the dependence with redshift is weak and using $R_{200c}/2$ still gives a major improvement. We leave a more detailed analysis of the redshift dependence of our results to a future work.
 


\section*{Acknowledgements}
We thank Subhabrata Majumdar, Romeel Dave, Nadia Zakamska, Shy Genel, Nick Battaglia, Luis Machado and Will Coulton for fruitful discussions. We also thank an anonymous referee for their critical comments as well as useful suggestions. We also thank Dylan Nelson, Annalisa Pillepich, Mark Vogelsberger for comments on the initial draft of this paper.
DW gratefully acknowledges support from the Friends of the Institute for Advanced Study Membership. JCH acknowledges support from NSF grant AST-2108536. FVN acknowledges funding from the WFIRST program through NNG26PJ30C and NNN12AA01C. DAA was supported in part by NSF grants AST-2009687 and AST-2108944. The work of SH is supported by Center for Computational Astrophysics of the Flatiron Institute in New York City. The Flatiron Institute is supported by the Simons Foundation.

\section*{Data Availability}

The code and data files associated with this paper are publicly available online \href{https://github.com/JayWadekar/ScalingRelations_ML}{\faGithub}.\footnote{\href{https://github.com/JayWadekar/ScalingRelations_ML}{\textcolor{blue}{https://github.com/JayWadekar/ScalingRelations\_ML}}.}

\bibliographystyle{mnras}
\bibliography{SZ}

\begin{thebibliography}{}
\makeatletter
\relax
\def\mn@urlcharsother{\let\do\@makeother \do\$\do\&\do\#\do\^\do\_\do\%\do\~}
\def\mn@doi{\begingroup\mn@urlcharsother \@ifnextchar [ {\mn@doi@}
  {\mn@doi@[]}}
\def\mn@doi@[#1]#2{\def\@tempa{#1}\ifx\@tempa\@empty \href
  {http://dx.doi.org/#2} {doi:#2}\else \href {http://dx.doi.org/#2} {#1}\fi
  \endgroup}
\def\mn@eprint#1#2{\mn@eprint@#1:#2::\@nil}
\def\mn@eprint@arXiv#1{\href {http://arxiv.org/abs/#1} {{\tt arXiv:#1}}}
\def\mn@eprint@dblp#1{\href {http://dblp.uni-trier.de/rec/bibtex/#1.xml}
  {dblp:#1}}
\def\mn@eprint@#1:#2:#3:#4\@nil{\def\@tempa {#1}\def\@tempb {#2}\def\@tempc
  {#3}\ifx \@tempc \@empty \let \@tempc \@tempb \let \@tempb \@tempa \fi \ifx
  \@tempb \@empty \def\@tempb {arXiv}\fi \@ifundefined
  {mn@eprint@\@tempb}{\@tempb:\@tempc}{\expandafter \expandafter \csname
  mn@eprint@\@tempb\endcsname \expandafter{\@tempc}}}

\bibitem[\protect\citeauthoryear{Ade et~al.}{Ade et~al.}{2016a}]{Ade15}
Ade P. A.~R.,  et~al., 2016a, \mn@doi [Astron. Astrophys.]
  {10.1051/0004-6361/201525833}, 594, A24

\bibitem[\protect\citeauthoryear{Ade et~al.}{Ade et~al.}{2016b}]{Ade15b}
Ade P. A.~R.,  et~al., 2016b, \mn@doi [Astron. Astrophys.]
  {10.1051/0004-6361/201525823}, 594, A27

\bibitem[\protect\citeauthoryear{Aghanim et~al.}{Aghanim et~al.}{2020}]{Aga18}
Aghanim N.,  et~al., 2020, \mn@doi [Astron. Astrophys.]
  {10.1051/0004-6361/201833910}, 641, A6

\bibitem[\protect\citeauthoryear{{Amodeo} et~al.}{{Amodeo}
  et~al.}{2021}]{Amo21}
{Amodeo} S.,  et~al., 2021, \mn@doi [\prd] {10.1103/PhysRevD.103.063514}, \href
  {https://ui.adsabs.harvard.edu/abs/2021PhRvD.103f3514A} {103, 063514}

\bibitem[\protect\citeauthoryear{{Amon}, {Gruen}  \& {DES
  Collaboration}}{{Amon} et~al.}{2022}]{Amo22}
{Amon} A.,  {Gruen} D.,   {DES Collaboration} 2022, \mn@doi [\prd]
  {10.1103/PhysRevD.105.023514}, \href
  {https://ui.adsabs.harvard.edu/abs/2022PhRvD.105b3514A} {105, 023514}

\bibitem[\protect\citeauthoryear{{Angl{\'e}s-Alc{\'a}zar}, {Dav{\'e}},
  {Faucher-Gigu{\`e}re}, {{\"O}zel}  \& {Hopkins}}{{Angl{\'e}s-Alc{\'a}zar}
  et~al.}{2017}]{Ang17}
{Angl{\'e}s-Alc{\'a}zar} D.,  {Dav{\'e}} R.,  {Faucher-Gigu{\`e}re} C.-A.,
  {{\"O}zel} F.,   {Hopkins} P.~F.,  2017, \mn@doi [\mnras]
  {10.1093/mnras/stw2565}, \href
  {https://ui.adsabs.harvard.edu/abs/2017MNRAS.464.2840A} {464, 2840}

\bibitem[\protect\citeauthoryear{{Arjona} \& {Nesseris}}{{Arjona} \&
  {Nesseris}}{2020}]{Arj20}
{Arjona} R.,  {Nesseris} S.,  2020, \mn@doi [\jcap]
  {10.1088/1475-7516/2020/11/042}, \href
  {https://ui.adsabs.harvard.edu/abs/2020JCAP...11..042A} {2020, 042}

\bibitem[\protect\citeauthoryear{{Avestruz}, {Lau}, {Nagai}  \&
  {Vikhlinin}}{{Avestruz} et~al.}{2014}]{Ave14}
{Avestruz} C.,  {Lau} E.~T.,  {Nagai} D.,   {Vikhlinin} A.,  2014, \mn@doi
  [\apj] {10.1088/0004-637X/791/2/117}, \href
  {https://ui.adsabs.harvard.edu/abs/2014ApJ...791..117A} {791, 117}

\bibitem[\protect\citeauthoryear{{Bartlett}, {Desmond}  \&
  {Ferreira}}{{Bartlett} et~al.}{2022}]{Bar22}
{Bartlett} D.~J.,  {Desmond} H.,   {Ferreira} P.~G.,  2022, arXiv e-prints,
  \href {https://ui.adsabs.harvard.edu/abs/2022arXiv221111461B} {p.
  arXiv:2211.11461}

\bibitem[\protect\citeauthoryear{{Battaglia}, {Bond}, {Pfrommer}  \&
  {Sievers}}{{Battaglia} et~al.}{2012a}]{BatBon12}
{Battaglia} N.,  {Bond} J.~R.,  {Pfrommer} C.,   {Sievers} J.~L.,  2012a,
  \mn@doi [\apj] {10.1088/0004-637X/758/2/74}, \href
  {https://ui.adsabs.harvard.edu/abs/2012ApJ...758...74B} {758, 74}

\bibitem[\protect\citeauthoryear{{Battaglia}, {Bond}, {Pfrommer}  \&
  {Sievers}}{{Battaglia} et~al.}{2012b}]{BatBon12b}
{Battaglia} N.,  {Bond} J.~R.,  {Pfrommer} C.,   {Sievers} J.~L.,  2012b,
  \mn@doi [\apj] {10.1088/0004-637X/758/2/75}, \href
  {https://ui.adsabs.harvard.edu/abs/2012ApJ...758...75B} {758, 75}

\bibitem[\protect\citeauthoryear{{Bayron Orjuela-Quintana}, {Nesseris}  \&
  {Cardona}}{{Bayron Orjuela-Quintana} et~al.}{2022}]{Bar23}
{Bayron Orjuela-Quintana} J.,  {Nesseris} S.,   {Cardona} W.,  2022, \mn@doi
  [arXiv e-prints] {10.48550/arXiv.2211.06393}, \href
  {https://ui.adsabs.harvard.edu/abs/2022arXiv221106393B} {p. arXiv:2211.06393}

\bibitem[\protect\citeauthoryear{{Bernal}, {Caputo}, {Villaescusa-Navarro}  \&
  {Kamionkowski}}{{Bernal} et~al.}{2021}]{Ber21}
{Bernal} J.~L.,  {Caputo} A.,  {Villaescusa-Navarro} F.,   {Kamionkowski} M.,
  2021, \mn@doi [\prl] {10.1103/PhysRevLett.127.131102}, \href
  {https://ui.adsabs.harvard.edu/abs/2021PhRvL.127m1102B} {127, 131102}

\bibitem[\protect\citeauthoryear{{Blanton} et~al.,}{{Blanton}
  et~al.}{2005}]{Bla05}
{Blanton} M.~R.,  et~al., 2005, \mn@doi [\aj] {10.1086/429803}, \href
  {https://ui.adsabs.harvard.edu/abs/2005AJ....129.2562B} {129, 2562}

\bibitem[\protect\citeauthoryear{Bocquet et~al.}{Bocquet et~al.}{2015}]{Boc15}
Bocquet S.,  et~al., 2015, \mn@doi [Astrophys. J.]
  {10.1088/0004-637X/799/2/214}, 799, 214

\bibitem[\protect\citeauthoryear{Bocquet et~al.}{Bocquet et~al.}{2019}]{Boc19}
Bocquet S.,  et~al., 2019, \mn@doi [Astrophys. J.] {10.3847/1538-4357/ab1f10},
  878, 55

\bibitem[\protect\citeauthoryear{{Booth} \& {Schaye}}{{Booth} \&
  {Schaye}}{2013}]{Boo13}
{Booth} C.~M.,  {Schaye} J.,  2013, \mn@doi [Scientific Reports]
  {10.1038/srep01738}, \href
  {https://ui.adsabs.harvard.edu/abs/2013NatSR....E1738B} {p.~1738}

\bibitem[\protect\citeauthoryear{{Bryan} \& {Norman}}{{Bryan} \&
  {Norman}}{1998}]{Bry98}
{Bryan} G.~L.,  {Norman} M.~L.,  1998, \mn@doi [\apj] {10.1086/305262}, \href
  {https://ui.adsabs.harvard.edu/abs/1998ApJ...495...80B} {495, 80}

\bibitem[\protect\citeauthoryear{{Chatterjee} \& {Kosowsky}}{{Chatterjee} \&
  {Kosowsky}}{2007}]{Cha07}
{Chatterjee} S.,  {Kosowsky} A.,  2007, \mn@doi [\apjl] {10.1086/518860}, \href
  {https://ui.adsabs.harvard.edu/abs/2007ApJ...661L.113C} {661, L113}

\bibitem[\protect\citeauthoryear{{Chatterjee}, {Di Matteo}, {Kosowsky}  \&
  {Pelupessy}}{{Chatterjee} et~al.}{2008}]{Cha08}
{Chatterjee} S.,  {Di Matteo} T.,  {Kosowsky} A.,   {Pelupessy} I.,  2008,
  \mn@doi [\mnras] {10.1111/j.1365-2966.2008.13784.x}, \href
  {https://ui.adsabs.harvard.edu/abs/2008MNRAS.390..535C} {390, 535}

\bibitem[\protect\citeauthoryear{{Chisari} et~al.,}{{Chisari}
  et~al.}{2018}]{Chi18}
{Chisari} N.~E.,  et~al., 2018, \mn@doi [\mnras] {10.1093/mnras/sty2093}, \href
  {https://ui.adsabs.harvard.edu/abs/2018MNRAS.480.3962C} {480, 3962}

\bibitem[\protect\citeauthoryear{{Chiu}, {Klein}, {Mohr}  \& {Bocquet}}{{Chiu}
  et~al.}{2022}]{Chi22}
{Chiu} I.-N.,  {Klein} M.,  {Mohr} J.,   {Bocquet} S.,  2022, arXiv e-prints,
  \href {https://ui.adsabs.harvard.edu/abs/2022arXiv220712429C} {p.
  arXiv:2207.12429}

\bibitem[\protect\citeauthoryear{Cranmer}{Cranmer}{2020}]{pysr}
Cranmer M.,  2020, PySR: Fast \& Parallelized Symbolic Regression in
  Python/Julia, \mn@doi{10.5281/zenodo.4052869}, \url
  {https://doi.org/10.5281/zenodo.4052869}

\bibitem[\protect\citeauthoryear{{Cranmer}, {Sanchez-Gonzalez}, {Battaglia},
  {Xu}, {Cranmer}, {Spergel}  \& {Ho}}{{Cranmer} et~al.}{2020}]{CraSan20}
{Cranmer} M.,  {Sanchez-Gonzalez} A.,  {Battaglia} P.,  {Xu} R.,  {Cranmer} K.,
   {Spergel} D.,   {Ho} S.,  2020, arXiv e-prints, \href
  {https://ui.adsabs.harvard.edu/abs/2020arXiv200611287C} {p. arXiv:2006.11287}

\bibitem[\protect\citeauthoryear{{Croston}, {Arnaud}, {Pointecouteau}  \&
  {Pratt}}{{Croston} et~al.}{2006}]{Cro06}
{Croston} J.~H.,  {Arnaud} M.,  {Pointecouteau} E.,   {Pratt} G.~W.,  2006,
  \mn@doi [\aap] {10.1051/0004-6361:20065795}, \href
  {https://ui.adsabs.harvard.edu/abs/2006A&A...459.1007C} {459, 1007}

\bibitem[\protect\citeauthoryear{{DESI Collaboration} et~al.}{{DESI
  Collaboration} et~al.}{2016}]{DESI}
{DESI Collaboration} et~al., 2016, arXiv e-prints, \href
  {https://ui.adsabs.harvard.edu/abs/2016arXiv161100036D} {p. arXiv:1611.00036}

\bibitem[\protect\citeauthoryear{{Dav{\'e}}, {Angl{\'e}s-Alc{\'a}zar},
  {Narayanan}, {Li}, {Rafieferantsoa}  \& {Appleby}}{{Dav{\'e}}
  et~al.}{2019}]{Dave2019}
{Dav{\'e}} R.,  {Angl{\'e}s-Alc{\'a}zar} D.,  {Narayanan} D.,  {Li} Q.,
  {Rafieferantsoa} M.~H.,   {Appleby} S.,  2019, \mn@doi [\mnras]
  {10.1093/mnras/stz937}, \href
  {https://ui.adsabs.harvard.edu/abs/2019MNRAS.486.2827D} {486, 2827}

\bibitem[\protect\citeauthoryear{{Delgado}, {Wadekar}, {Hadzhiyska}, {Bose},
  {Hernquist}  \& {Ho}}{{Delgado} et~al.}{2021}]{Del21}
{Delgado} A.~M.,  {Wadekar} D.,  {Hadzhiyska} B.,  {Bose} S.,  {Hernquist} L.,
   {Ho} S.,  2021, arXiv e-prints, \href
  {https://ui.adsabs.harvard.edu/abs/2021arXiv211102422D} {p. arXiv:2111.02422}

\bibitem[\protect\citeauthoryear{{Dutta Chowdhury} \& {Chatterjee}}{{Dutta
  Chowdhury} \& {Chatterjee}}{2017}]{Dut17}
{Dutta Chowdhury} D.,  {Chatterjee} S.,  2017, \mn@doi [\apj]
  {10.3847/1538-4357/aa64d6}, \href
  {https://ui.adsabs.harvard.edu/abs/2017ApJ...839...34D} {839, 34}

\bibitem[\protect\citeauthoryear{{Gatti} et~al.}{{Gatti} et~al.}{2021}]{Gat21}
{Gatti} M.,  et~al., 2021, arXiv e-prints, \href
  {https://ui.adsabs.harvard.edu/abs/2021arXiv210801600G} {p. arXiv:2108.01600}

\bibitem[\protect\citeauthoryear{{Graham}, {Djorgovski}, {Mahabal}, {Donalek},
  {Drake}  \& {Longo}}{{Graham} et~al.}{2012}]{astroexample3}
{Graham} M.~J.,  {Djorgovski} S.~G.,  {Mahabal} A.,  {Donalek} C.,  {Drake} A.,
    {Longo} G.,  2012, arXiv e-prints, \href
  {https://ui.adsabs.harvard.edu/abs/2012arXiv1208.2480G} {p. arXiv:1208.2480}

\bibitem[\protect\citeauthoryear{{Graham}, {Djorgovski}, {Mahabal}, {Donalek}
  \& {Drake}}{{Graham} et~al.}{2013}]{astroexample2}
{Graham} M.~J.,  {Djorgovski} S.~G.,  {Mahabal} A.~A.,  {Donalek} C.,   {Drake}
  A.~J.,  2013, \mn@doi [\mnras] {10.1093/mnras/stt329}, \href
  {https://ui.adsabs.harvard.edu/abs/2013MNRAS.431.2371G} {431, 2371}

\bibitem[\protect\citeauthoryear{{Greco}, {Hill}, {Spergel}  \&
  {Battaglia}}{{Greco} et~al.}{2015}]{Gre15}
{Greco} J.~P.,  {Hill} J.~C.,  {Spergel} D.~N.,   {Battaglia} N.,  2015,
  \mn@doi [\apj] {10.1088/0004-637X/808/2/151}, \href
  {https://ui.adsabs.harvard.edu/abs/2015ApJ...808..151G} {808, 151}

\bibitem[\protect\citeauthoryear{{Greene}, {Strader}  \& {Ho}}{{Greene}
  et~al.}{2020}]{Gre20}
{Greene} J.~E.,  {Strader} J.,   {Ho} L.~C.,  2020, \mn@doi [\araa]
  {10.1146/annurev-astro-032620-021835}, \href
  {https://ui.adsabs.harvard.edu/abs/2020ARA&A..58..257G} {58, 257}

\bibitem[\protect\citeauthoryear{{Hahn} et~al.,}{{Hahn} et~al.}{2022}]{Hah22}
{Hahn} C.,  et~al., 2022, arXiv e-prints, \href
  {https://ui.adsabs.harvard.edu/abs/2022arXiv220201809H} {p. arXiv:2202.01809}

\bibitem[\protect\citeauthoryear{{Hall} et~al.,}{{Hall} et~al.}{2019}]{Hal19}
{Hall} K.~R.,  et~al., 2019, \mn@doi [\mnras] {10.1093/mnras/stz2751}, \href
  {https://ui.adsabs.harvard.edu/abs/2019MNRAS.490.2315H} {490, 2315}

\bibitem[\protect\citeauthoryear{{Hand} et~al.}{{Hand} et~al.}{2011}]{Han11}
{Hand} N.,  et~al., 2011, \mn@doi [\apj] {10.1088/0004-637X/736/1/39}, \href
  {https://ui.adsabs.harvard.edu/abs/2011ApJ...736...39H} {736, 39}

\bibitem[\protect\citeauthoryear{{Harnois-D{\'e}raps}, {van Waerbeke}, {Viola}
  \& {Heymans}}{{Harnois-D{\'e}raps} et~al.}{2015}]{Har15}
{Harnois-D{\'e}raps} J.,  {van Waerbeke} L.,  {Viola} M.,   {Heymans} C.,
  2015, \mn@doi [\mnras] {10.1093/mnras/stv646}, \href
  {https://ui.adsabs.harvard.edu/abs/2015MNRAS.450.1212H} {450, 1212}

\bibitem[\protect\citeauthoryear{{Hasselfield} et~al.}{{Hasselfield}
  et~al.}{2013}]{Has13}
{Hasselfield} M.,  et~al., 2013, \mn@doi [\jcap]
  {10.1088/1475-7516/2013/07/008}, \href
  {https://ui.adsabs.harvard.edu/abs/2013JCAP...07..008H} {2013, 008}

\bibitem[\protect\citeauthoryear{{Hill}, {Baxter}, {Lidz}, {Greco}  \&
  {Jain}}{{Hill} et~al.}{2018}]{Hil18}
{Hill} J.~C.,  {Baxter} E.~J.,  {Lidz} A.,  {Greco} J.~P.,   {Jain} B.,  2018,
  \mn@doi [\prd] {10.1103/PhysRevD.97.083501}, \href
  {https://ui.adsabs.harvard.edu/abs/2018PhRvD..97h3501H} {97, 083501}

\bibitem[\protect\citeauthoryear{{Hilton} et~al.}{{Hilton}
  et~al.}{2021}]{Hil21}
{Hilton} M.,  et~al., 2021, \mn@doi [\apjs] {10.3847/1538-4365/abd023}, \href
  {https://ui.adsabs.harvard.edu/abs/2021ApJS..253....3H} {253, 3}

\bibitem[\protect\citeauthoryear{{Hopkins}}{{Hopkins}}{2015}]{Hop15}
{Hopkins} P.~F.,  2015, \mn@doi [\mnras] {10.1093/mnras/stv195}, \href
  {https://ui.adsabs.harvard.edu/abs/2015MNRAS.450...53H} {450, 53}

\bibitem[\protect\citeauthoryear{{Hopkins}}{{Hopkins}}{2017}]{Hop17}
{Hopkins} P.~F.,  2017, arXiv e-prints, \href
  {https://ui.adsabs.harvard.edu/abs/2017arXiv171201294H} {p. arXiv:1712.01294}

\bibitem[\protect\citeauthoryear{{Jimeno}, {Diego}, {Broadhurst}, {De Martino}
  \& {Lazkoz}}{{Jimeno} et~al.}{2018}]{Jim18}
{Jimeno} P.,  {Diego} J.~M.,  {Broadhurst} T.,  {De Martino} I.,   {Lazkoz} R.,
   2018, \mn@doi [\mnras] {10.1093/mnras/sty987}, \href
  {https://ui.adsabs.harvard.edu/abs/2018MNRAS.478..638J} {478, 638}

\bibitem[\protect\citeauthoryear{{Kaiser}}{{Kaiser}}{1986}]{Kai86}
{Kaiser} N.,  1986, \mn@doi [\mnras] {10.1093/mnras/222.2.323}, \href
  {https://ui.adsabs.harvard.edu/abs/1986MNRAS.222..323K} {222, 323}

\bibitem[\protect\citeauthoryear{{Kravtsov} \& {Borgani}}{{Kravtsov} \&
  {Borgani}}{2012}]{KraBor12}
{Kravtsov} A.~V.,  {Borgani} S.,  2012, \mn@doi [\araa]
  {10.1146/annurev-astro-081811-125502}, \href
  {https://ui.adsabs.harvard.edu/abs/2012ARA&A..50..353K} {50, 353}

\bibitem[\protect\citeauthoryear{{Kravtsov}, {Nagai}  \&
  {Vikhlinin}}{{Kravtsov} et~al.}{2005}]{Kra05}
{Kravtsov} A.~V.,  {Nagai} D.,   {Vikhlinin} A.~A.,  2005, \mn@doi [\apj]
  {10.1086/429796}, \href
  {https://ui.adsabs.harvard.edu/abs/2005ApJ...625..588K} {625, 588}

\bibitem[\protect\citeauthoryear{{LSST Dark Energy Science
  Collaboration}}{{LSST Dark Energy Science Collaboration}}{2012}]{Rubin}
{LSST Dark Energy Science Collaboration} 2012, arXiv e-prints, \href
  {https://ui.adsabs.harvard.edu/abs/2012arXiv1211.0310L} {p. arXiv:1211.0310}

\bibitem[\protect\citeauthoryear{{Lacy} et~al.,}{{Lacy} et~al.}{2019}]{Lac19}
{Lacy} M.,  et~al., 2019, \mn@doi [\mnras] {10.1093/mnrasl/sly215}, \href
  {https://ui.adsabs.harvard.edu/abs/2019MNRAS.483L..22L} {483, L22}

\bibitem[\protect\citeauthoryear{{Le Brun}, {McCarthy}  \& {Melin}}{{Le Brun}
  et~al.}{2015}]{LeB15}
{Le Brun} A. M.~C.,  {McCarthy} I.~G.,   {Melin} J.-B.,  2015, \mn@doi [\mnras]
  {10.1093/mnras/stv1172}, \href
  {https://ui.adsabs.harvard.edu/abs/2015MNRAS.451.3868L} {451, 3868}

\bibitem[\protect\citeauthoryear{{Le Brun}, {McCarthy}, {Schaye}  \&
  {Ponman}}{{Le Brun} et~al.}{2017}]{LeB17}
{Le Brun} A. M.~C.,  {McCarthy} I.~G.,  {Schaye} J.,   {Ponman} T.~J.,  2017,
  \mn@doi [\mnras] {10.1093/mnras/stw3361}, \href
  {https://ui.adsabs.harvard.edu/abs/2017MNRAS.466.4442L} {466, 4442}

\bibitem[\protect\citeauthoryear{{Lee}, {Coulton}, {Thiele}  \& {Ho}}{{Lee}
  et~al.}{2022}]{Lee22}
{Lee} B. K.~K.,  {Coulton} W.~R.,  {Thiele} L.,   {Ho} S.,  2022, arXiv
  e-prints, \href {https://ui.adsabs.harvard.edu/abs/2022arXiv220501710L} {p.
  arXiv:2205.01710}

\bibitem[\protect\citeauthoryear{{Lemos}, {Jeffrey}, {Cranmer}, {Ho}  \&
  {Battaglia}}{{Lemos} et~al.}{2022}]{Lem22}
{Lemos} P.,  {Jeffrey} N.,  {Cranmer} M.,  {Ho} S.,   {Battaglia} P.,  2022,
  arXiv e-prints, \href {https://ui.adsabs.harvard.edu/abs/2022arXiv220202306L}
  {p. arXiv:2202.02306}

\bibitem[\protect\citeauthoryear{{Liu} et~al.}{{Liu} et~al.}{2022}]{Liu22}
{Liu} A.,  et~al., 2022, \mn@doi [\aap] {10.1051/0004-6361/202141120}, \href
  {https://ui.adsabs.harvard.edu/abs/2022A&A...661A...2L} {661, A2}

\bibitem[\protect\citeauthoryear{{Marinacci} et~al.,}{{Marinacci}
  et~al.}{2018}]{Mar18}
{Marinacci} F.,  et~al., 2018, \mn@doi [\mnras] {10.1093/mnras/sty2206}, \href
  {https://ui.adsabs.harvard.edu/abs/2018MNRAS.480.5113M} {480, 5113}

\bibitem[\protect\citeauthoryear{{Moser} et~al.,}{{Moser} et~al.}{2022}]{Mos22}
{Moser} E.,  et~al., 2022, \mn@doi [\apj] {10.3847/1538-4357/ac70c6}, \href
  {https://ui.adsabs.harvard.edu/abs/2022ApJ...933..133M} {933, 133}

\bibitem[\protect\citeauthoryear{{Nagai} \& {Lau}}{{Nagai} \&
  {Lau}}{2011}]{Nag11}
{Nagai} D.,  {Lau} E.~T.,  2011, \mn@doi [\apjl] {10.1088/2041-8205/731/1/L10},
  \href {https://ui.adsabs.harvard.edu/abs/2011ApJ...731L..10N} {731, L10}

\bibitem[\protect\citeauthoryear{{Naiman} et~al.,}{{Naiman}
  et~al.}{2018}]{Nai18}
{Naiman} J.~P.,  et~al., 2018, \mn@doi [\mnras] {10.1093/mnras/sty618}, \href
  {https://ui.adsabs.harvard.edu/abs/2018MNRAS.477.1206N} {477, 1206}

\bibitem[\protect\citeauthoryear{{Nelson} et~al.,}{{Nelson}
  et~al.}{2018}]{Nel18}
{Nelson} D.,  et~al., 2018, \mn@doi [\mnras] {10.1093/mnras/stx3040}, \href
  {https://ui.adsabs.harvard.edu/abs/2018MNRAS.475..624N} {475, 624}

\bibitem[\protect\citeauthoryear{{Nelson} et~al.,}{{Nelson}
  et~al.}{2019}]{Nel19}
{Nelson} D.,  et~al., 2019, \mn@doi [Computational Astrophysics and Cosmology]
  {10.1186/s40668-019-0028-x}, \href
  {https://ui.adsabs.harvard.edu/abs/2019ComAC...6....2N} {6, 2}

\bibitem[\protect\citeauthoryear{{Nicola} et~al.,}{{Nicola}
  et~al.}{2022}]{Nic22}
{Nicola} A.,  et~al., 2022, \mn@doi [\jcap] {10.1088/1475-7516/2022/04/046},
  \href {https://ui.adsabs.harvard.edu/abs/2022JCAP...04..046N} {2022, 046}

\bibitem[\protect\citeauthoryear{{Osato}, {Flender}, {Nagai}, {Shirasaki}  \&
  {Yoshida}}{{Osato} et~al.}{2018}]{Osa18}
{Osato} K.,  {Flender} S.,  {Nagai} D.,  {Shirasaki} M.,   {Yoshida} N.,  2018,
  \mn@doi [\mnras] {10.1093/mnras/stx3215}, \href
  {https://ui.adsabs.harvard.edu/abs/2018MNRAS.475..532O} {475, 532}

\bibitem[\protect\citeauthoryear{{Osato}, {Shirasaki}, {Miyatake}, {Nagai},
  {Yoshida}, {Oguri}  \& {Takahashi}}{{Osato} et~al.}{2020}]{Osa20}
{Osato} K.,  {Shirasaki} M.,  {Miyatake} H.,  {Nagai} D.,  {Yoshida} N.,
  {Oguri} M.,   {Takahashi} R.,  2020, \mn@doi [\mnras]
  {10.1093/mnras/staa117}, \href
  {https://ui.adsabs.harvard.edu/abs/2020MNRAS.492.4780O} {492, 4780}

\bibitem[\protect\citeauthoryear{{Palmese} et~al.}{{Palmese}
  et~al.}{2020}]{Pal20}
{Palmese} A.,  et~al., 2020, \mn@doi [\mnras] {10.1093/mnras/staa526}, \href
  {https://ui.adsabs.harvard.edu/abs/2020MNRAS.493.4591P} {493, 4591}

\bibitem[\protect\citeauthoryear{{Pandey}, {Baxter}  \& {Hill}}{{Pandey}
  et~al.}{2020}]{Pan20}
{Pandey} S.,  {Baxter} E.~J.,   {Hill} J.~C.,  2020, \mn@doi [\prd]
  {10.1103/PhysRevD.101.043525}, \href
  {https://ui.adsabs.harvard.edu/abs/2020PhRvD.101d3525P} {101, 043525}

\bibitem[\protect\citeauthoryear{{Pandey} et~al.}{{Pandey}
  et~al.}{2022}]{Pan21}
{Pandey} S.,  et~al., 2022, \mn@doi [\prd] {10.1103/PhysRevD.105.123526}, \href
  {https://ui.adsabs.harvard.edu/abs/2022PhRvD.105l3526P} {105, 123526}

\bibitem[\protect\citeauthoryear{{Pillepich} et~al.,}{{Pillepich}
  et~al.}{2018a}]{PilSprNel1801}
{Pillepich} A.,  et~al., 2018a, \mn@doi [\mnras] {10.1093/mnras/stx2656}, \href
  {https://ui.adsabs.harvard.edu/abs/2018MNRAS.473.4077P} {473, 4077}

\bibitem[\protect\citeauthoryear{{Pillepich} et~al.,}{{Pillepich}
  et~al.}{2018b}]{Pil18}
{Pillepich} A.,  et~al., 2018b, \mn@doi [\mnras] {10.1093/mnras/stx3112}, \href
  {https://ui.adsabs.harvard.edu/abs/2018MNRAS.475..648P} {475, 648}

\bibitem[\protect\citeauthoryear{{Planck Collaboration} et~al.}{{Planck
  Collaboration} et~al.}{2011}]{Aga11}
{Planck Collaboration} et~al., 2011, \mn@doi [\aap]
  {10.1051/0004-6361/201116489}, \href
  {https://ui.adsabs.harvard.edu/abs/2011A&A...536A..12P} {536, A12}

\bibitem[\protect\citeauthoryear{{Planck Collaboration}, {Ade}  et~al.}{{Planck
  Collaboration} et~al.}{2013}]{Ade13}
{Planck Collaboration} {Ade} P.~A.~R.,   et~al., 2013, \mn@doi [\aap]
  {10.1051/0004-6361/201220941}, \href
  {https://ui.adsabs.harvard.edu/abs/2013A&A...557A..52P} {557, A52}

\bibitem[\protect\citeauthoryear{{Pop} et~al.,}{{Pop} et~al.}{2022}]{Pop21}
{Pop} A.-R.,  et~al., 2022, arXiv e-prints, \href
  {https://ui.adsabs.harvard.edu/abs/2022arXiv220511528P} {p. arXiv:2205.11528}

\bibitem[\protect\citeauthoryear{{Robson} \& {Dav{\'e}}}{{Robson} \&
  {Dav{\'e}}}{2020}]{RobDav20}
{Robson} D.,  {Dav{\'e}} R.,  2020, \mn@doi [\mnras] {10.1093/mnras/staa2394},
  \href {https://ui.adsabs.harvard.edu/abs/2020MNRAS.498.3061R} {498, 3061}

\bibitem[\protect\citeauthoryear{{Robson} \& {Dav{\'e}}}{{Robson} \&
  {Dav{\'e}}}{2021}]{RobDav21}
{Robson} D.,  {Dav{\'e}} R.,  2021, arXiv e-prints, \href
  {https://ui.adsabs.harvard.edu/abs/2021arXiv210701206R} {p. arXiv:2107.01206}

\bibitem[\protect\citeauthoryear{{Ruan}, {McQuinn}  \& {Anderson}}{{Ruan}
  et~al.}{2015}]{Rua15}
{Ruan} J.~J.,  {McQuinn} M.,   {Anderson} S.~F.,  2015, \mn@doi [\apj]
  {10.1088/0004-637X/802/2/135}, \href
  {https://ui.adsabs.harvard.edu/abs/2015ApJ...802..135R} {802, 135}

\bibitem[\protect\citeauthoryear{{Schaan} \& {Atacama Cosmology Telescope
  Collaboration}}{{Schaan} \& {Atacama Cosmology Telescope
  Collaboration}}{2021}]{Sch21}
{Schaan} E.,  {Atacama Cosmology Telescope Collaboration} 2021, \mn@doi [\prd]
  {10.1103/PhysRevD.103.063513}, \href
  {https://ui.adsabs.harvard.edu/abs/2021PhRvD.103f3513S} {103, 063513}

\bibitem[\protect\citeauthoryear{{Secco}, {Samuroff}  \& {DES
  Collaboration}}{{Secco} et~al.}{2022}]{Sec22}
{Secco} L.~F.,  {Samuroff} S.,   {DES Collaboration} 2022, \mn@doi [\prd]
  {10.1103/PhysRevD.105.023515}, \href
  {https://ui.adsabs.harvard.edu/abs/2022PhRvD.105b3515S} {105, 023515}

\bibitem[\protect\citeauthoryear{{Semboloni}, {Hoekstra}, {Schaye}, {van
  Daalen}  \& {McCarthy}}{{Semboloni} et~al.}{2011}]{Sem11}
{Semboloni} E.,  {Hoekstra} H.,  {Schaye} J.,  {van Daalen} M.~P.,   {McCarthy}
  I.~G.,  2011, \mn@doi [\mnras] {10.1111/j.1365-2966.2011.19385.x}, \href
  {https://ui.adsabs.harvard.edu/abs/2011MNRAS.417.2020S} {417, 2020}

\bibitem[\protect\citeauthoryear{{Shao} et~al.,}{{Shao} et~al.}{2022}]{Sha21}
{Shao} H.,  et~al., 2022, \mn@doi [\apj] {10.3847/1538-4357/ac4d30}, \href
  {https://ui.adsabs.harvard.edu/abs/2022ApJ...927...85S} {927, 85}

\bibitem[\protect\citeauthoryear{{Shao} et~al.,}{{Shao} et~al.}{2023}]{Sha23}
{Shao} H.,  et~al., 2023, \mn@doi [arXiv e-prints] {10.48550/arXiv.2302.14591},
  \href {https://ui.adsabs.harvard.edu/abs/2023arXiv230214591S} {p.
  arXiv:2302.14591}

\bibitem[\protect\citeauthoryear{{Singh}, {Voit}  \& {Nath}}{{Singh}
  et~al.}{2021}]{Sin21}
{Singh} P.,  {Voit} G.~M.,   {Nath} B.~B.,  2021, \mn@doi [\mnras]
  {10.1093/mnras/staa3827}, \href
  {https://ui.adsabs.harvard.edu/abs/2021MNRAS.501.2467S} {501, 2467}

\bibitem[\protect\citeauthoryear{{Soergel}, {Giannantonio}, {Efstathiou},
  {Puchwein}  \& {Sijacki}}{{Soergel} et~al.}{2017}]{Soe17}
{Soergel} B.,  {Giannantonio} T.,  {Efstathiou} G.,  {Puchwein} E.,   {Sijacki}
  D.,  2017, \mn@doi [\mnras] {10.1093/mnras/stx492}, \href
  {https://ui.adsabs.harvard.edu/abs/2017MNRAS.468..577S} {468, 577}

\bibitem[\protect\citeauthoryear{{Somerville} \& {Dav{\'e}}}{{Somerville} \&
  {Dav{\'e}}}{2015}]{Som15}
{Somerville} R.~S.,  {Dav{\'e}} R.,  2015, \mn@doi [\araa]
  {10.1146/annurev-astro-082812-140951}, \href
  {https://ui.adsabs.harvard.edu/abs/2015ARA&A..53...51S} {53, 51}

\bibitem[\protect\citeauthoryear{{Springel}}{{Springel}}{2010}]{Spr10}
{Springel} V.,  2010, \mn@doi [\mnras] {10.1111/j.1365-2966.2009.15715.x},
  \href {https://ui.adsabs.harvard.edu/abs/2010MNRAS.401..791S} {401, 791}

\bibitem[\protect\citeauthoryear{{Springel} et~al.,}{{Springel}
  et~al.}{2018}]{Spr18}
{Springel} V.,  et~al., 2018, \mn@doi [\mnras] {10.1093/mnras/stx3304}, \href
  {https://ui.adsabs.harvard.edu/abs/2018MNRAS.475..676S} {475, 676}

\bibitem[\protect\citeauthoryear{{Sun}, {Voit}, {Donahue}, {Jones}, {Forman}
  \& {Vikhlinin}}{{Sun} et~al.}{2009}]{Sun09}
{Sun} M.,  {Voit} G.~M.,  {Donahue} M.,  {Jones} C.,  {Forman} W.,
  {Vikhlinin} A.,  2009, \mn@doi [\apj] {10.1088/0004-637X/693/2/1142}, \href
  {https://ui.adsabs.harvard.edu/abs/2009ApJ...693.1142S} {693, 1142}

\bibitem[\protect\citeauthoryear{{Tegmark}}{{Tegmark}}{1997}]{Teg9711}
{Tegmark} M.,  1997, \mn@doi [\prl] {10.1103/PhysRevLett.79.3806}, \href
  {https://ui.adsabs.harvard.edu/abs/1997PhRvL..79.3806T} {79, 3806}

\bibitem[\protect\citeauthoryear{{Thiele} et~al.,}{{Thiele}
  et~al.}{2022}]{Thi22}
{Thiele} L.,  et~al., 2022, \mn@doi [\prd] {10.1103/PhysRevD.105.083505}, \href
  {https://ui.adsabs.harvard.edu/abs/2022PhRvD.105h3505T} {105, 083505}

\bibitem[\protect\citeauthoryear{{Tinker}, {Kravtsov}, {Klypin}, {Abazajian},
  {Warren}, {Yepes}, {Gottl{\"o}ber}  \& {Holz}}{{Tinker} et~al.}{2008}]{Tin08}
{Tinker} J.,  {Kravtsov} A.~V.,  {Klypin} A.,  {Abazajian} K.,  {Warren} M.,
  {Yepes} G.,  {Gottl{\"o}ber} S.,   {Holz} D.~E.,  2008, \mn@doi [\apj]
  {10.1086/591439}, \href
  {https://ui.adsabs.harvard.edu/abs/2008ApJ...688..709T} {688, 709}

\bibitem[\protect\citeauthoryear{{Van Daalen}, {Schaye}, {McCarthy}, {Booth}
  \& {Dalla Vecchia}}{{Van Daalen} et~al.}{2014}]{Van14}
{Van Daalen} M.~P.,  {Schaye} J.,  {McCarthy} I.~G.,  {Booth} C.~M.,   {Dalla
  Vecchia} C.,  2014, \mn@doi [\mnras] {10.1093/mnras/stu482}, \href
  {https://ui.adsabs.harvard.edu/abs/2014MNRAS.440.2997V} {440, 2997}

\bibitem[\protect\citeauthoryear{{Villaescusa-Navarro}
  et~al.,}{{Villaescusa-Navarro} et~al.}{2021}]{VilAngGen20}
{Villaescusa-Navarro} F.,  et~al., 2021, \mn@doi [\apj]
  {10.3847/1538-4357/abf7ba}, \href
  {https://ui.adsabs.harvard.edu/abs/2021ApJ...915...71V} {915, 71}

\bibitem[\protect\citeauthoryear{{Villaescusa-Navarro}
  et~al.,}{{Villaescusa-Navarro} et~al.}{2022}]{CAMELS_public}
{Villaescusa-Navarro} F.,  et~al., 2022, arXiv e-prints, \href
  {https://ui.adsabs.harvard.edu/abs/2022arXiv220101300V} {p. arXiv:2201.01300}

\bibitem[\protect\citeauthoryear{{Voevodkin} \& {Vikhlinin}}{{Voevodkin} \&
  {Vikhlinin}}{2004}]{Voe04}
{Voevodkin} A.,  {Vikhlinin} A.,  2004, \mn@doi [\apj] {10.1086/380818}, \href
  {https://ui.adsabs.harvard.edu/abs/2004ApJ...601..610V} {601, 610}

\bibitem[\protect\citeauthoryear{{Vogelsberger} et~al.,}{{Vogelsberger}
  et~al.}{2014a}]{Vog14}
{Vogelsberger} M.,  et~al., 2014a, \mn@doi [\mnras] {10.1093/mnras/stu1536},
  \href {https://ui.adsabs.harvard.edu/abs/2014MNRAS.444.1518V} {444, 1518}

\bibitem[\protect\citeauthoryear{{Vogelsberger} et~al.,}{{Vogelsberger}
  et~al.}{2014b}]{Vog14b}
{Vogelsberger} M.,  et~al., 2014b, \mn@doi [\nat] {10.1038/nature13316}, \href
  {https://ui.adsabs.harvard.edu/abs/2014Natur.509..177V} {509, 177}

\bibitem[\protect\citeauthoryear{{Vogelsberger}, {Marinacci}, {Torrey}  \&
  {Puchwein}}{{Vogelsberger} et~al.}{2020}]{Vog20}
{Vogelsberger} M.,  {Marinacci} F.,  {Torrey} P.,   {Puchwein} E.,  2020,
  \mn@doi [Nature Reviews Physics] {10.1038/s42254-019-0127-2}, \href
  {https://ui.adsabs.harvard.edu/abs/2020NatRP...2...42V} {2, 42}

\bibitem[\protect\citeauthoryear{{Wadekar}, {Villaescusa-Navarro}, {Ho}  \&
  {Perreault-Levasseur}}{{Wadekar} et~al.}{2020a}]{WadVil20b}
{Wadekar} D.,  {Villaescusa-Navarro} F.,  {Ho} S.,   {Perreault-Levasseur} L.,
  2020a, arXiv e-prints, \href
  {https://ui.adsabs.harvard.edu/abs/2020arXiv201200111W} {p. arXiv:2012.00111}

\bibitem[\protect\citeauthoryear{{Wadekar}, {Ivanov}  \&
  {Scoccimarro}}{{Wadekar} et~al.}{2020b}]{WadIva20}
{Wadekar} D.,  {Ivanov} M.~M.,   {Scoccimarro} R.,  2020b, \mn@doi [\prd]
  {10.1103/PhysRevD.102.123521}, \href
  {https://ui.adsabs.harvard.edu/abs/2020PhRvD.102l3521W} {102, 123521}

\bibitem[\protect\citeauthoryear{{Wadekar} et~al.,}{{Wadekar}
  et~al.}{2022}]{Wad22}
{Wadekar} D.,  et~al., 2022, arXiv e-prints, \href
  {https://ui.adsabs.harvard.edu/abs/2022arXiv220101305W} {p. arXiv:2201.01305}

\bibitem[\protect\citeauthoryear{{Wechsler} \& {Tinker}}{{Wechsler} \&
  {Tinker}}{2018}]{Wec18}
{Wechsler} R.~H.,  {Tinker} J.~L.,  2018, \mn@doi [\araa]
  {10.1146/annurev-astro-081817-051756}, \href
  {https://ui.adsabs.harvard.edu/abs/2018ARA&A..56..435W} {56, 435}

\bibitem[\protect\citeauthoryear{{Weinberger} et~al.,}{{Weinberger}
  et~al.}{2017}]{Wei17}
{Weinberger} R.,  et~al., 2017, \mn@doi [\mnras] {10.1093/mnras/stw2944}, \href
  {https://ui.adsabs.harvard.edu/abs/2017MNRAS.465.3291W} {465, 3291}

\bibitem[\protect\citeauthoryear{{Weinberger}, {Springel}  \&
  {Pakmor}}{{Weinberger} et~al.}{2020}]{Wei20}
{Weinberger} R.,  {Springel} V.,   {Pakmor} R.,  2020, \mn@doi [\apjs]
  {10.3847/1538-4365/ab908c}, \href
  {https://ui.adsabs.harvard.edu/abs/2020ApJS..248...32W} {248, 32}

\bibitem[\protect\citeauthoryear{{Wong} \& {Cranmer}}{{Wong} \&
  {Cranmer}}{2022}]{Won22}
{Wong} K. W.~K.,  {Cranmer} M.,  2022, arXiv e-prints, \href
  {https://ui.adsabs.harvard.edu/abs/2022arXiv220712409W} {p. arXiv:2207.12409}

\bibitem[\protect\citeauthoryear{{Yang}, {Cai}, {Cui}, {Dav{\'e}}, {Peacock}
  \& {Sorini}}{{Yang} et~al.}{2022}]{Yan22}
{Yang} T.,  {Cai} Y.-C.,  {Cui} W.,  {Dav{\'e}} R.,  {Peacock} J.~A.,
  {Sorini} D.,  2022, arXiv e-prints, \href
  {https://ui.adsabs.harvard.edu/abs/2022arXiv220211430Y} {p. arXiv:2202.11430}

\makeatother
\end{thebibliography}

\appendix


\section{Supplementary analysis}
\label{apx:details}

\subsection{Additional results from the random forest}
\label{apx:RF}
In section~\ref{sec:Results1}, we had trained the random forest (RF) on gas and stellar mass related properties. We now probe whether adding two additional properties can further improve the RF performance. 

\begin{figure}
\centering
\includegraphics[scale=0.55,keepaspectratio=true]{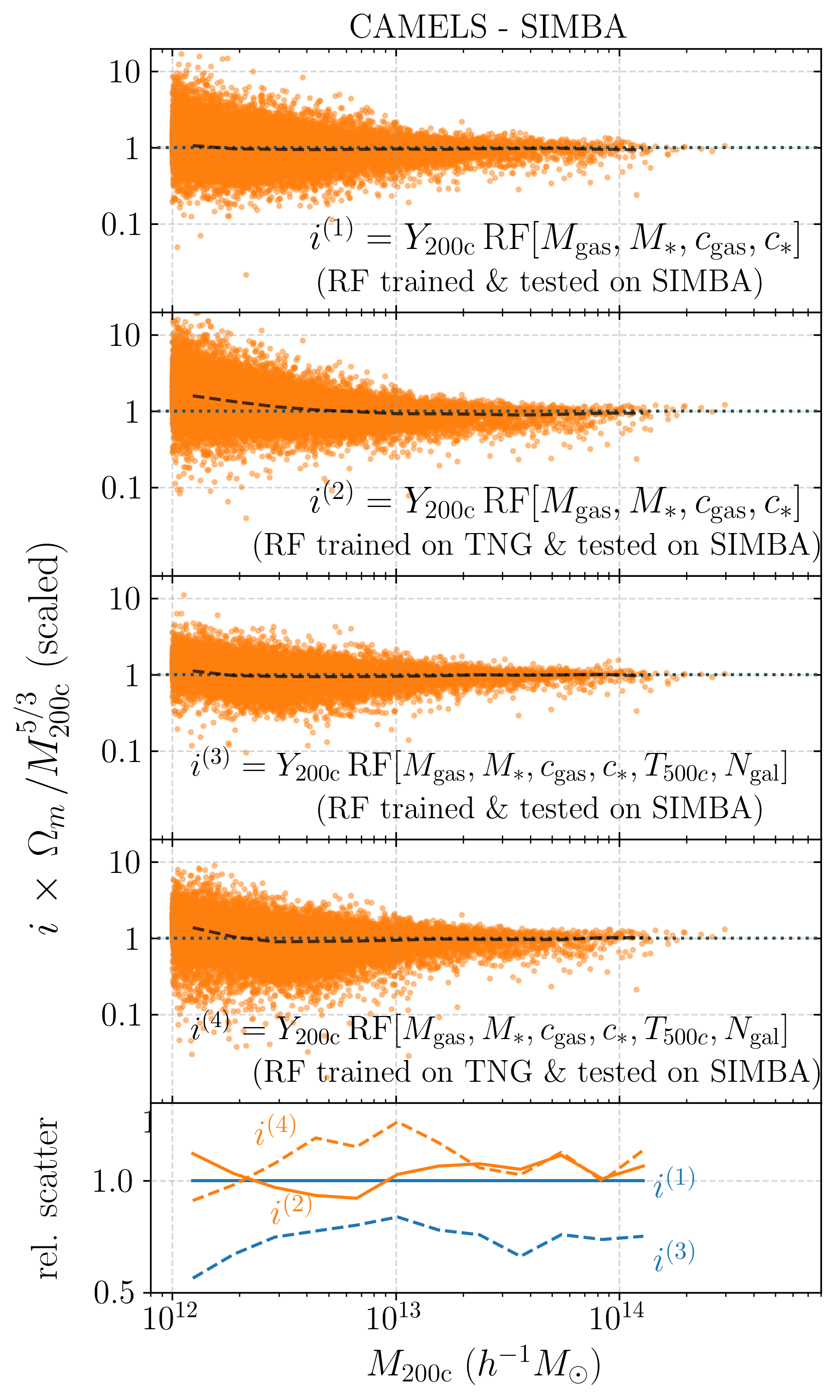}
\includegraphics[scale=0.55,keepaspectratio=true]{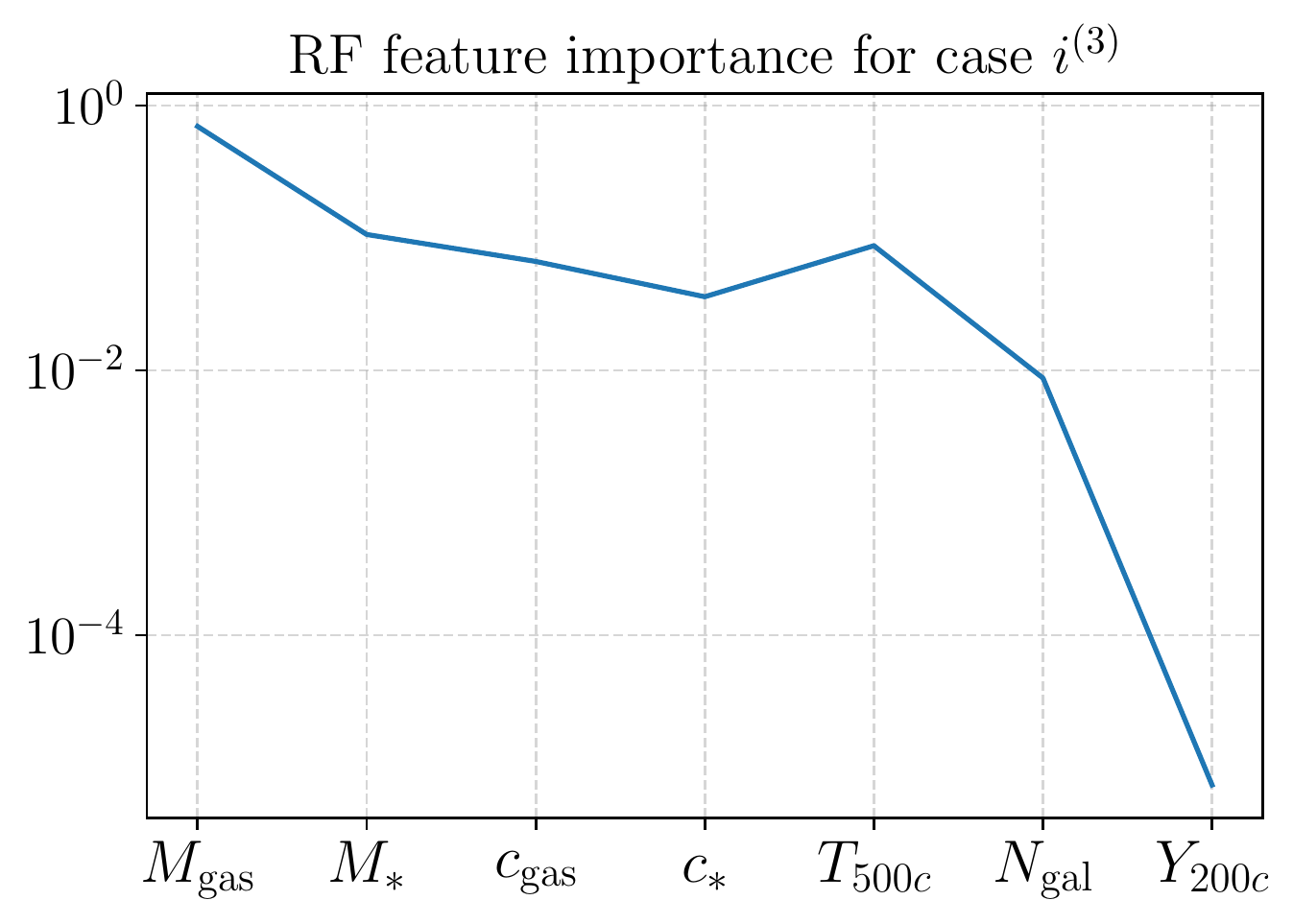}
\caption{\textbf{Top}: Same as Fig.~\ref{fig:RF} but showing results from the RF with two additional parameters: average temperature of gas within $R_{500c}$ ($T_{500c}$), and the number of galaxies associated with the halo ($N_\mathrm{gal}$). \textbf{Bottom}: The importance of different input variables for the random forest (RF) prediction in the case $i^{(3)}$.} 
\label{fig:FeatureImp}
\end{figure}

The first property we consider is $T_{500c}$ which is the average temperature of hot gas within $R_{500c}$. Our definition of hot gas is similar to \cite{Yan22} and it refers to the particles with $T>10^5$ K (as both hydrogen and helium are present in ionized state). The second property is the total number of galaxies ($N_\mathrm{gal}$) associated with the halos which have $M_*>10^9\Ms$. We show results in Fig.~\ref{fig:FeatureImp}. We see that including $T_{500c}$ and $N_\mathrm{gal}$ does improve the RF prediction when both trained and tested on SIMBA halos. However, the improvement does not directly generalize when the RF trained on TNG halos and tested on SIMBA halos. This could be because the correlation of $T_{500c}$ and $N_\mathrm{gal}$ with the $Y-M$ relation could be different for the case of TNG and SIMBA, and one might therefore need to introduce additional fitting parameters to properly model their dependence.

\begin{figure}
\centering
\includegraphics[scale=0.55,keepaspectratio=true]{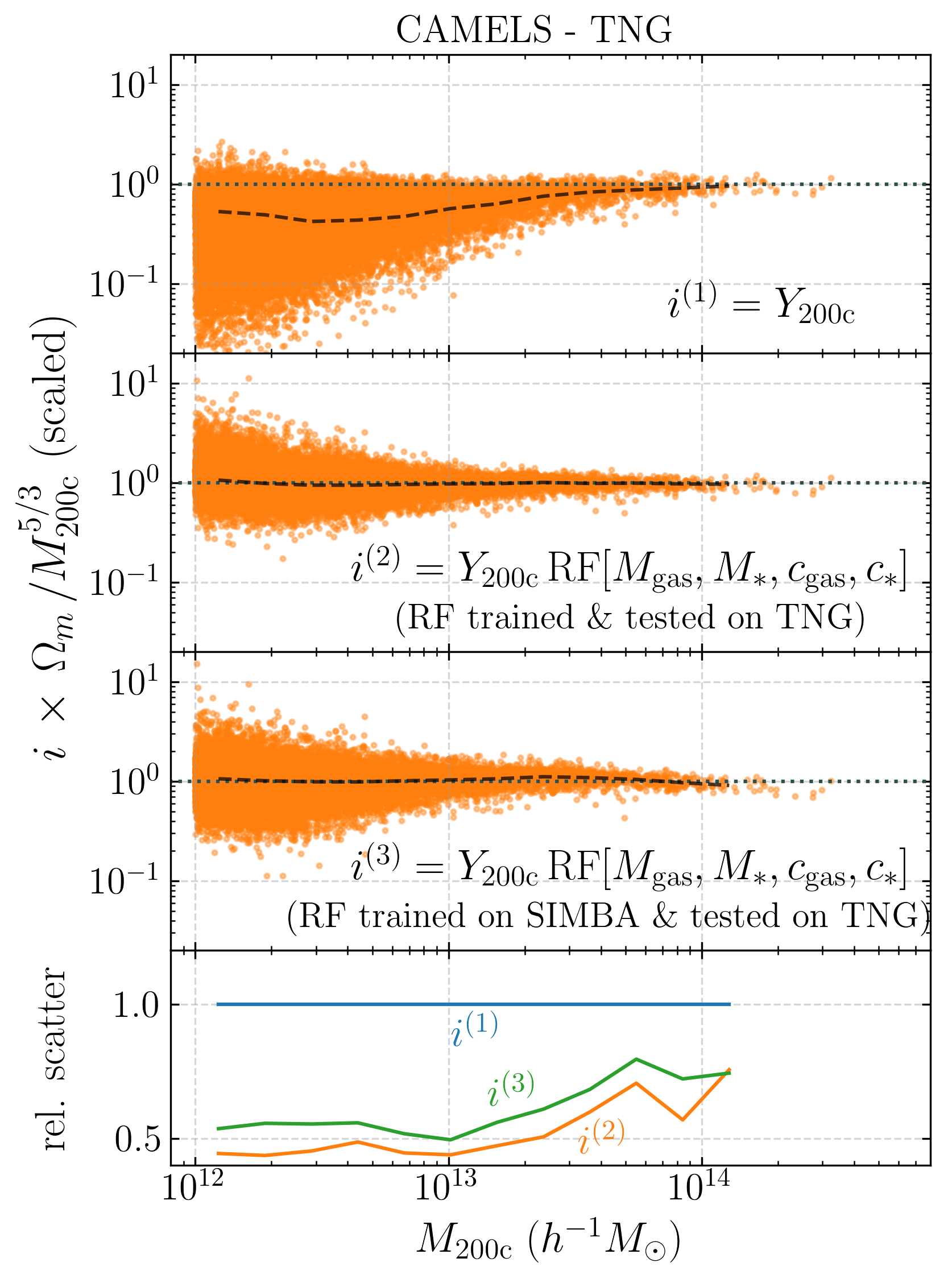}
\caption{Same as Fig.~\ref{fig:RF}, but for testing the RF with halos from CAMELS-TNG simulations instead of CAMELS-SIMBA ones. Again, the RF augmented predictions have a substantially smaller deviation from a power-law relation, and the scatter is also smaller.}
\label{fig:RF_TNG}
\end{figure}

 We can also obtain the relative importance of input features once the RF is trained and we show this data in the bottom panel of Fig.~\ref{fig:FeatureImp}. Overall, $T_{500c}$ could be an important parameter as it is correlated with the deviation from self-similarity in the low-mass $Y-M$ relation. We also show a plot similar to Fig.~\ref{fig:RF} but for CAMELS-TNG halos in Fig.~\ref{fig:RF_TNG}.

\subsection{Additional results from symbolic regression and possible improvements}
\label{apx:SR}
\begin{figure*}
\centering
\includegraphics[scale=0.55,keepaspectratio=true]{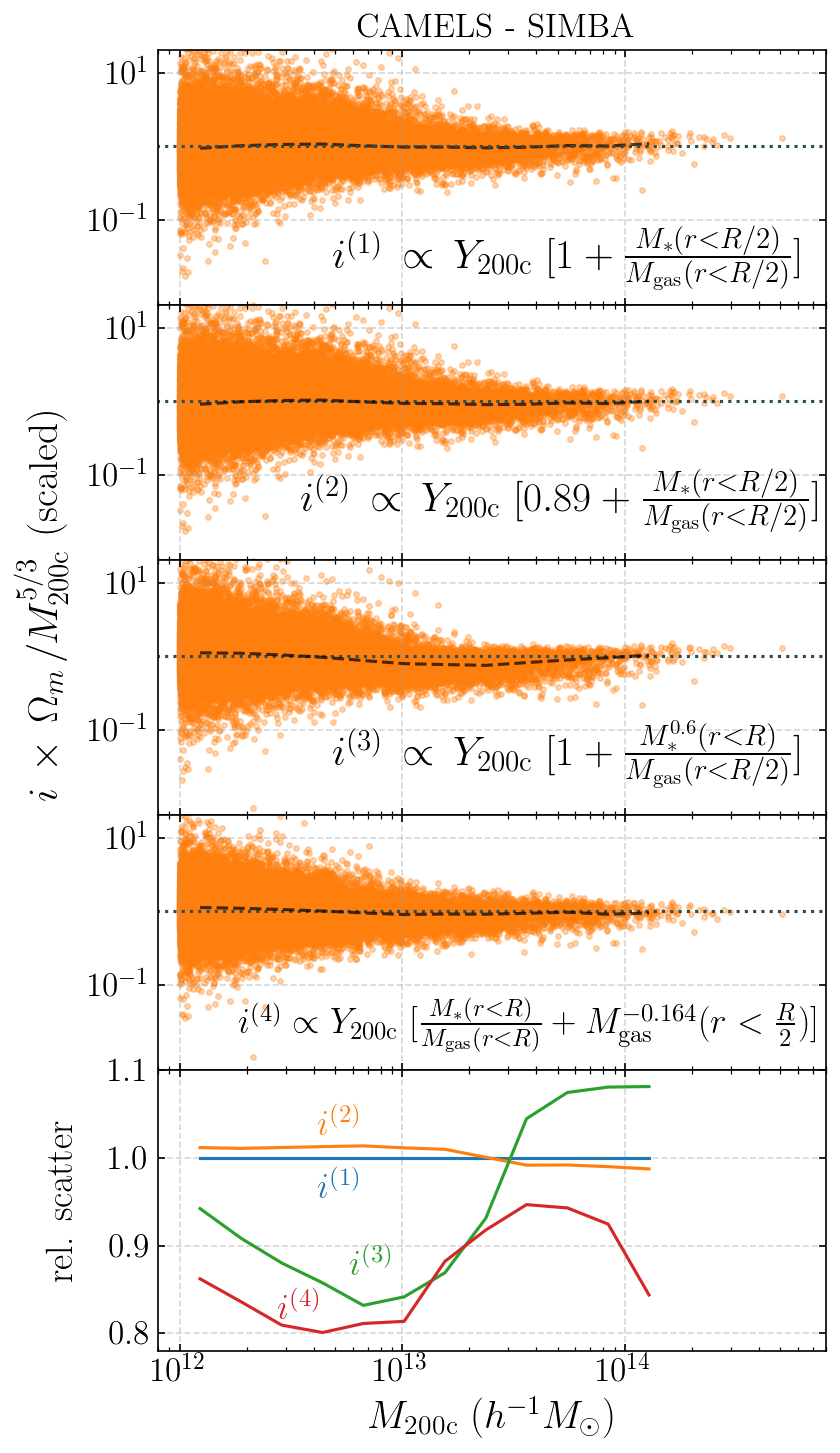}
\includegraphics[scale=0.55,keepaspectratio=true]{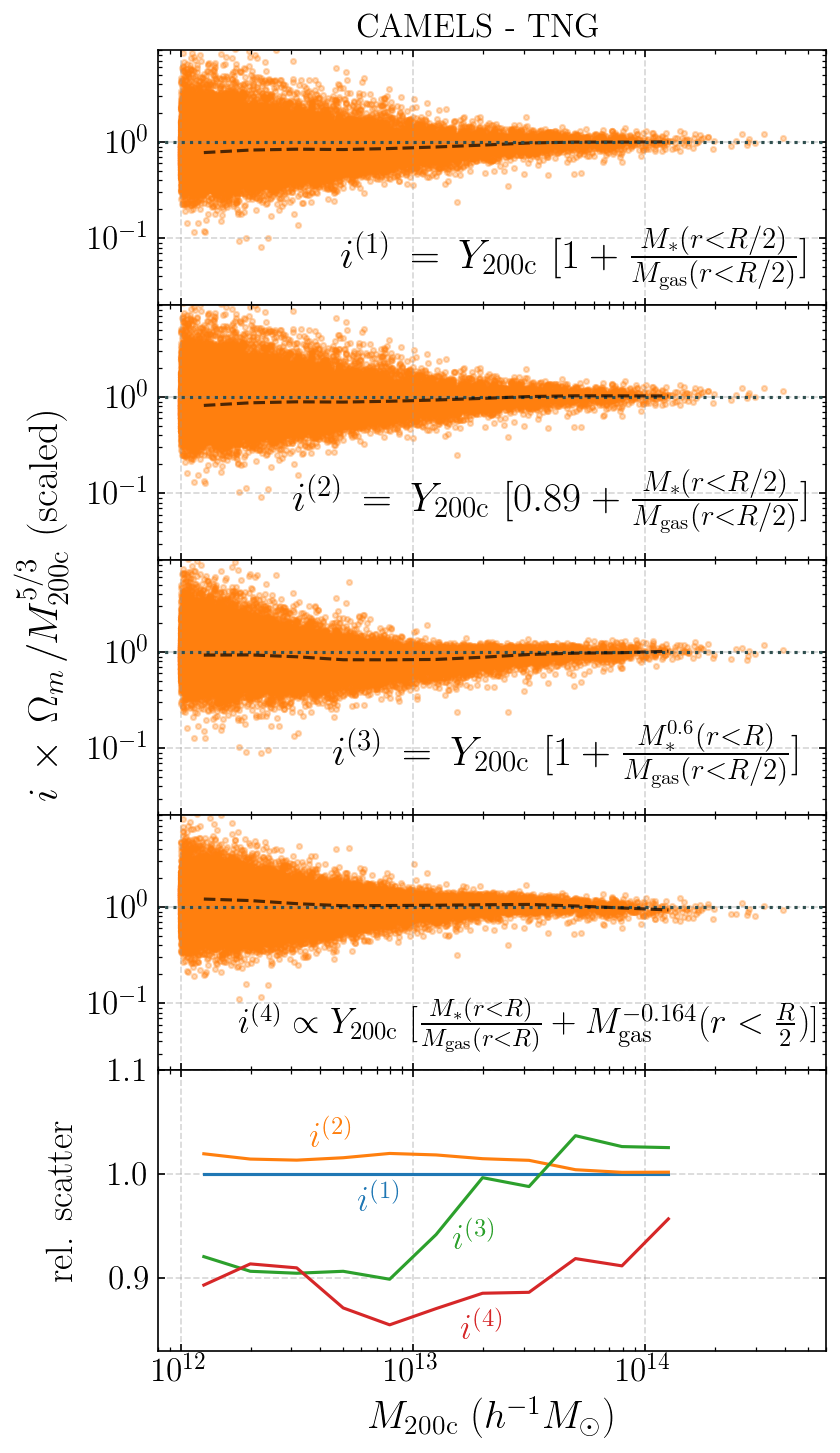}
\caption{Same as Fig.~\ref{fig:Y_Mgal}, but showing here three additional results from symbolic regression (the top panel contains the best result from Fig.~\ref{fig:Y_Mgal} and is included here for comparison). Note that $M_*$ and $M_\mathrm{gas}$ in the equations are normalized by $10^{10}\Ms$.}
\label{fig:Y_Mgal2}
\end{figure*}

We showed a few results obtained from SR in Fig.~\ref{fig:Y_Mgal} of the main text, here we show some additional results in Fig.~\ref{fig:Y_Mgal2}. We see that more complex equations (see e.g., the $i^{(4)}$ relation in the figure) can fit the data better (and also generalize for the case of different sub-grid presciptions), but they have more complex forms.

We only included $\{M_* (R_{200c})$, $M_\mathrm{gas}(R_{200c}),M_*(R_{200c}/2)$, $M_\mathrm{gas}(R_{200c}/2)\}$ as the input set to the SR in this paper. It would be interesting to run SR including the parameters $T_{500c}$ and $N_\mathrm{gal}$ in the input set, but we leave that to a future analysis. 

In Fig.~\ref{fig:Y_Mgal2}, we show the equations with exactly the same fitting coefficients for both TNG and SIMBA. It would be interesting to tweak the fitting coefficients (but keeping the equation form to be the same) separately for the two cases. Currently, we constucted the input set for SR by combining halos from both TNG and SIMBA (so the same equation and coefficients were obtained by design). It would be interesting to train the SR on halos from either one, and then to tweak the fitting parameters of the resulting equations to best fit the other case.

\begin{figure*}
\centering
\includegraphics[scale=0.5,keepaspectratio=true]{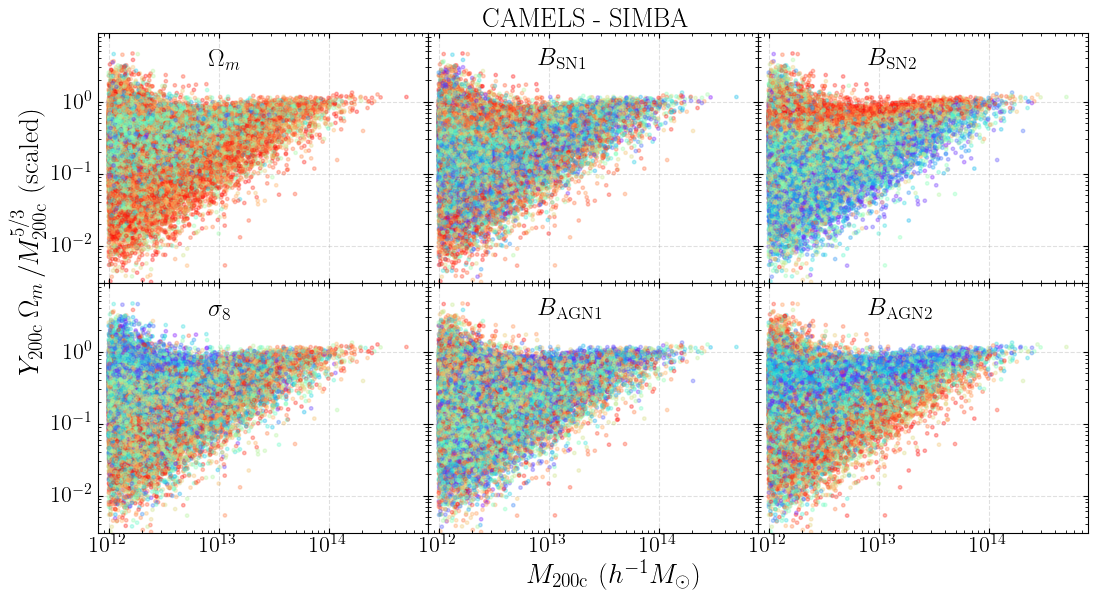}
\includegraphics[scale=0.25,keepaspectratio=true]{Figs/colorbar.pdf}
\includegraphics[scale=0.5,keepaspectratio=true]{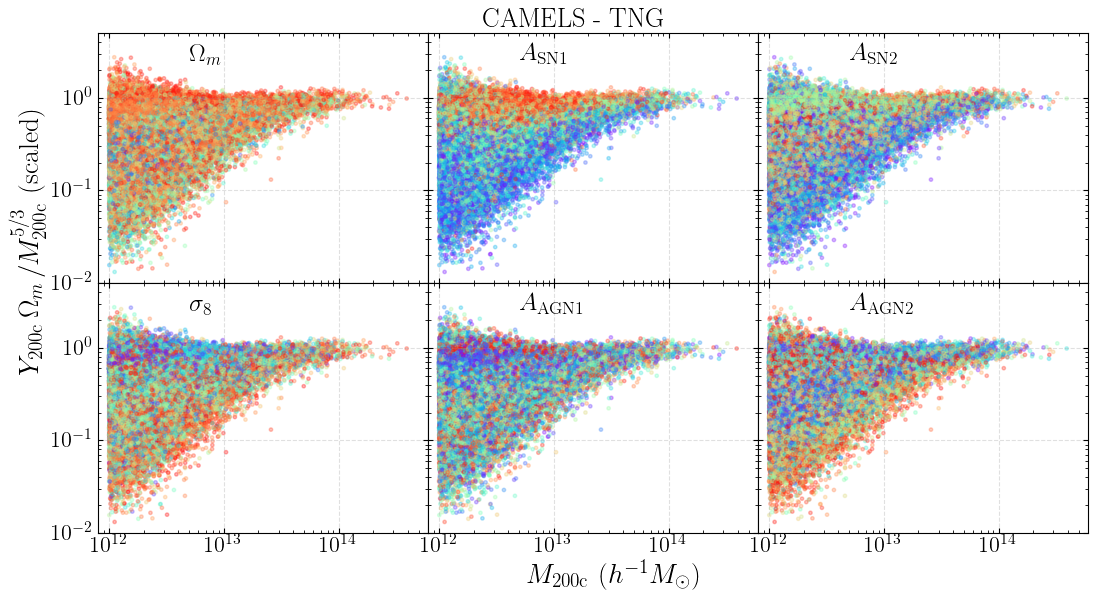}
\includegraphics[scale=0.25,keepaspectratio=true]{Figs/blank.pdf}
\caption{Same as Figure~\ref{fig:AstroProps1P}, but when all the six parameters are simultaneously varied in a latin hypercube (LH) fashion.
Unlike the 1P set, the LH set also includes variations in cosmic seeds of the simulation, and contains 1000 simulations each for TNG and SIMBA. It is interesting to note that the effects of SN and AGN feedback parameters are of roughly similar magnitudes on the $Y-M$ relation.} 
\label{fig:AstroPropsLH}
\end{figure*}

\begin{figure*}
\centering
\includegraphics[scale=0.55,keepaspectratio=true]{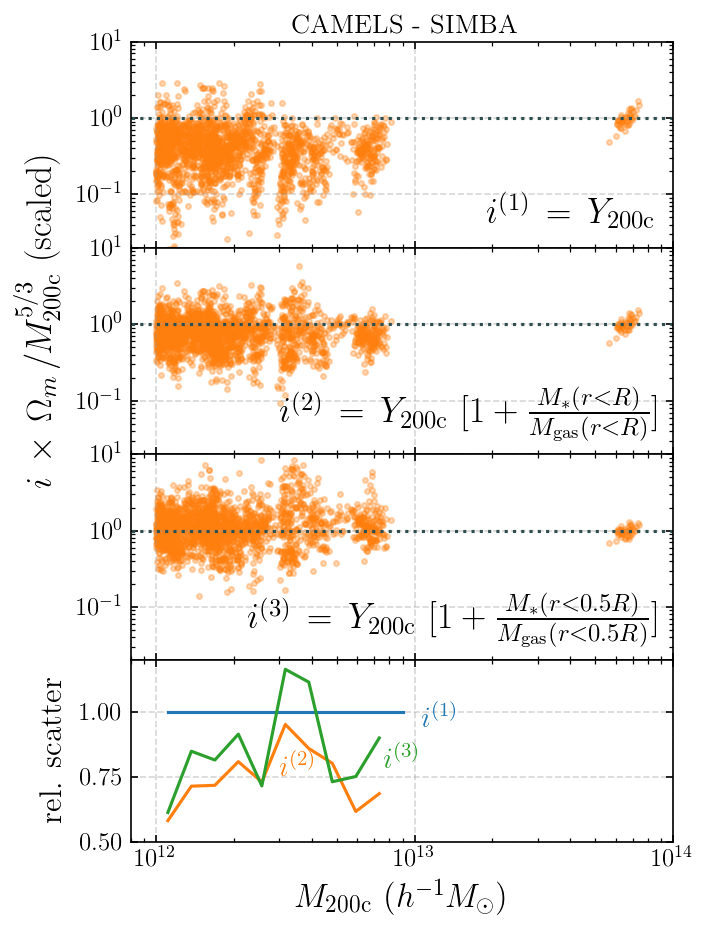}
\includegraphics[scale=0.55,keepaspectratio=true]{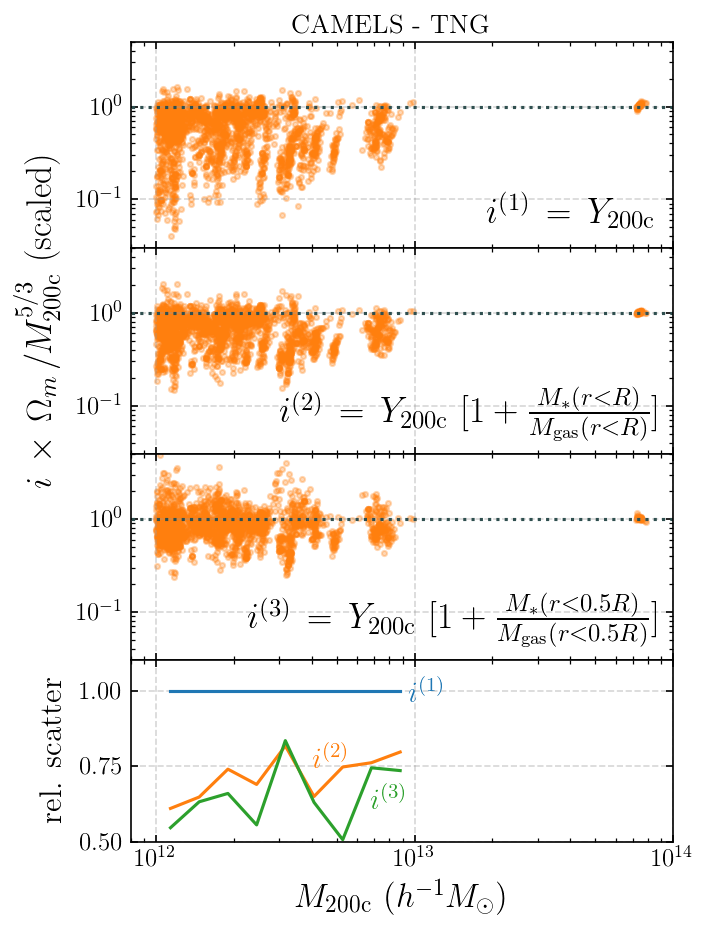}
\caption{Same as Fig.~\ref{fig:Y_Mgal}, but comparing results for the 1P set for only the cases where the four feedback parameters (SN1, SN2, AGN1, AGN2) are varied (keeping $\Omega_m, \sigma_8$ fixed to their fiducial values of 0.3 and 0.8 respectively). We again find similar level of improvement. Note that, compared to Fig.~\ref{fig:Y_Mgal}, there is additional stochasticity in the calculation of scatter in this figure due to the lower number of simulations used (44 simulations  as compared to 1000 simulations for Fig.~\ref{fig:Y_Mgal}).}
\label{fig:Y_Mgal_1P}
\end{figure*}

\bsp	
\label{lastpage}
\end{document}